\documentclass[journal]{IEEEtran}
\usepackage{graphicx}
\usepackage[table,xcdraw]{xcolor}
\usepackage{amsmath}
\usepackage{amssymb}
\usepackage{booktabs}
\usepackage{multirow}
\usepackage{color}
\usepackage{adjustbox}
\usepackage{xcolor}
\usepackage{cite}
\usepackage{float}
\usepackage{dblfloatfix}
\usepackage{algorithm}
\usepackage[vlined, ruled, algo2e]{algorithm2e}
\usepackage{caption}
\usepackage{subcaption}
\usepackage{csquotes}
\usepackage{float}

\usepackage[inline]{enumitem}
\usepackage{varwidth}% http://ctan.org/pkg/varwidth
\ifodd 1
\newcommand*\rot{\rotatebox{90}}
\else

\fi
\begin{document}
\title{The Study of Urban Residential's Public Space Activeness using Space-Centric Approach}

\author{Billy~Pik~Lik~Lau,~\IEEEmembership{Graduate Student Member,~IEEE,}
		Benny~Kai~Kiat~Ng,
        Chau~Yuen,~\IEEEmembership{Fellow,~IEEE,}
        Bige~Tun\c{c}er,
 		and Keng~Hua~Chong%
\thanks{Billy Pik Lik Lau, Benny Kai Kiat Ng, and Chau Yuen are with the Engineering Product Development, Singapore University of Technology and Design, Corresponding E-mail:
	 billy\_lau@mymail.sutd.edu.sg, yuenchau@sutd.edu.sg.}% 
\thanks{Prof. Bige Tun\c{c}er and Keng~Hua~Chong are with the Architecture and Sustainable Design, Singapore University of Technology and Design.}}
%\thanks{Manuscript received Ja/n 11, 2020}}

\markboth{Accepted at IEEE Internet of Things Journal, 2021}%
{Shell \MakeLowercase{\textit{et al.}}: Bare Demo of IEEEtran.cls for IEEE Journals}

\maketitle

\begin{abstract}
With the advancement of the Internet of Things (IoT) and communication platform, large scale sensor deployment can be easily implemented in an urban city to collect various information.
To date, there are only a handful of research studies about understanding the usage of urban public spaces.
Leveraging IoT, various sensors have been deployed in an urban residential area to monitor and study public space utilization patterns.
In this paper, we propose a data processing system to generate space-centric insights about the utilization of an urban residential region of multiple points of interest (PoIs) that consists of 190,000m$^2$ real estate.
We identify the activeness of each PoI based on the spectral clustering, and then study their corresponding static features, which are composed of transportation, commercial facilities, population density, along with other characteristics.
Through the heuristic features inferring, the residential density and commercial facilities are the most significant factors affecting public place utilization.
\end{abstract}

\begin{IEEEkeywords}
Space-centric Monitoring, Public Space Utilization, Spatial Temporal, Internet of Things, Smart City
\end{IEEEkeywords}

\IEEEpeerreviewmaketitle

%%%%%%%%%%%%%%%%%%%%%%%%%%%%%%%%%%%%%%%%%%%%%%%%%%%%%%%%%%%%%%%%%%%%%%%%%%%%%%%%%%%%%%%%%%%%%%%%%%%%%%%%%%
\section{Introduction}
\label{sec:Introduction}
%%%%%%%%%%%%%%%%%%%%%%%%%%%%%%%%%%%%%%%%%%%%%%%%%%%%%%%%%%%%%%%%%%%%%%%%%%%%%%%%%%%%%%%%%%%%%%%%%%%%%%%%%%
\IEEEPARstart{T}{he} advancement of communication technologies and low-cost sensors have paved the direction for the Internet of things (IoT) to be widely deployed on a large scale, which is currently one of the most prominent research areas in Smart City development.
It allows a government agency to monitor various regions of the city for different information such as traffic management, sewer management, smart building, etc.
The advancement of low power devices and communication technology has granted the possibility of pervasive monitoring for a long period of duration in a smart city. 
To date, various smart city applications have gained attention especially in improving various aspects of the smart city as reviewed in~\cite{Zanella2014Internet} and~\cite{lau2019survey}, which consists of medical devices~\cite{Connor2017Privacy,solanas2014smart}, waste management~\cite{anagnostopoulos2017challenges}, traffic monitoring~\cite{wang2015understanding}, smart energy systems~\cite{spano2015last,mckenna2015four}, pervasive environmental monitoring~\cite{bacco2017environmental,lau2016spatial,Gamanayake2020Cluster}, and miscellaneous.

Among them, human behavior in an urban city is one of the most discussed topics by city governors as it often influences the planning decision of the urban development.
Urban planners often face the difficulty of planning a more livable area for the citizen and often it requires different kinds of data sources to evaluate before commencing a new project to build or renovate public spaces.
One of the common sensor devices that most urban citizens carry is a smartphone, and it can be used to capture data related to human behavior as shown in~\cite{Liu2018Survey,Marakkalage2019Understanding,Koh2020Multiple}.
However, the ease of collecting private data causes privacy intrusion issues and therefore only limited information is available to understand the urban citizen's behavior.
This can be an obstacle for organizations that have the real intention of studying human behavior data rather than exploiting it for certain benefits.
Despite users' consent being given, one potential cyber threat is information theft, which can be discouraging for urban citizens participating in studies about human behavior.
Data collection methods without directly involving users have been used to address the privacy intrusion problem such as passive WiFi sniffer~\cite{Petre2017Chapter,zhou2020understanding}, sonar sensors\cite{li2015senseflow,viswanath2014design}, motion sensors\cite{lau2016spatial, lau2018sensor}, etc.

Upon further investigation with these approaches, we notice that the majority of the literature consists of people-centric monitoring techniques, which track people over a specific time-line using the physical sensor and extract insights from it.
When considering the aforementioned privacy intrusion problem, there arises a need for a space-centric monitoring approach to eliminate privacy intrusion problems.
Examples of such methodology can be found in~\cite{junior2010crowd} and~\cite{bacco2017environmental}. 
To date, most space-centric methods have been focusing on the indoor environment due to various challenges such as limited coverage, weatherproof hardware, and other types of challenges.
Thus, it motivates us to study the space-centric monitoring of an urban residential area in an outdoor environment.

In this paper, we propose a space-centric approach architecture to monitor public space utilization in an urban residential area and study the activeness of outdoor public spaces.
Subsequently, the potential factors that affect the activeness of outdoor public spaces were studied.
Hence, the data pipeline is proposed to process the utilization data acquired by the space-centric sensors and extract the underlying features by grouping them based on the activeness of PoI.
Next, the potential factors that affect the activeness of a location can be conducted by correlating the activeness of PoI with the static features, which are composed of transportation, commercial facilities, population density, along with other characteristics.

The contributions in this paper are three-fold and listed as follows:
\begin{itemize}
	\item A data pipeline model is proposed to profile data collected from multiple outdoor space-centric sensors spread across an urban region.
	\item The activeness of a PoI for the studied region is defined by clustering their utilization profiles on three different types of day, which are weekday, weekend, and school holiday.
	\item The potential factors that drive the activeness of public space were analyzed heuristically based on static features.
\end{itemize}

The remaining paper can be organized into the following sections:
First, we discuss related works regarding space-centric monitoring in Section~\ref{sec:relatedWorks}.
Next, the data pipeline for processing different types of data collected from PoIs in the designated region is introduced in Section~\ref{sec:dataPipeline}.
Afterward, in Section~\ref{sec:sensorProfiling}, we briefly discuss the underlying mechanism for profiling the sensors along with the similarity kernel.
Using the aforementioned sensor profiling, the utilization pattern for each PoI is studied in Section~\ref{sec:magnitudeUtz} along with their static features.
Lastly, we conclude our findings in Section~\ref{sec:conclusion}.

%%%%%%%%%%%%%%%%%%%%%%%%%%%%%%%%%%%%%%%%%%%%%%%%%%%%%%%%%%%%%%%%%%%%%%%%%%%%%%%%%%%%%%%%%%%%%%%%%%%%%%%%%%
\section{Related Works on Space-Centric Monitoring}
\label{sec:relatedWorks}
%%%%%%%%%%%%%%%%%%%%%%%%%%%%%%%%%%%%%%%%%%%%%%%%%%%%%%%%%%%%%%%%%%%%%%%%%%%%%%%%%%%%%%%%%%%%%%%%%%%%%%%%%%
In this section, we present the literature works that are related to the space-centric monitoring of human activity. 
Specifically, data sources and data extraction techniques are emphasized.

\vspace{-0.28cm}

\subsection{Space Centric Data Sources}
The data sources for space-centric monitoring can be generally divided into three different categories such as physical data sources (obtained from sensors), cyber-data sources (using on-line resources), and participation (crowd survey).

Physical data sources involve actual sensors to collect data at a particular location over a certain period.
For instance, motion~\cite{raykov2016predicting,lau2016spatial} and proximity sensors~\cite{viswanath2014design,Choi2018Bi} are frequently used to monitor a particular space of interest.
These sensing techniques normally focus on one PoI due to limited coverage, and therefore deployment of multiple sensor units is often required to cover large areas.
Also, computer vision is common in tracking the activeness of a particular space of interest using cameras~\cite{junior2010crowd,chan2008privacy,Xing2011Robust,Gamanayake2020Cluster} and thermal imaging~\cite{tyndall2016occupancy,Amin2008Automated}.
However, it requires high computation and large data storage space, which is not ideal for long term deployment using sustainable energy sources.
Moreover, sound data can be used in space-centric monitoring as shown in~\cite{barchiesi2015acoustic,salamon2015unsupervised,Sanchez2017Using}.
It offers monitoring by understanding the acoustic characteristics of the surveillance of the specific area, but it is sensitive to a noisy environment and requires a high sampling rate. 
It makes it less ideal to deploy long-term, and the ground truth for the data modeling often requires a lot of human labor.
Likewise, building management systems as shown in~\cite{candanedo2016accurate,akbar2015contextual,Kleiminger2015Household,tushar2018internet} in frequently leverage space-centric data as input data sources.
They use electrical consumption as a metric to measure the occupancy of a place, but this has limited coverage up to a single building only.
As for cyber-data and participatory data, there is only a handful of literature to date.
For instance, cyber-data such as geotagged data as shown in~\cite{Birenboim2015High} as well as~\cite{FernandezVilas2019Analysis} is mainly used in space-centric data extraction for analyzing the activeness of a particular public space. 
An example of the participatory data can be found in~\cite{Chong2019Role} and~\cite{Chong2019When}, which explores the opinions of residents, and attempts to understand the usage of public space.

After surveying the related work about space-centric data sources, we notice the majority of the data sources focus on the indoor environment. 
An indoor environment is a space that is easier to set up when compared to the outdoor environment.
This introduces a research gap in understanding outdoor public space.

\vspace{-0.28cm}

\subsection{Data Extraction Techniques}
To understand insights and knowledge about a particular space of interest, various platforms and techniques are proposed. 
In this subsection, the data extraction techniques can be categorized into the following groups: state estimation/inference and machine learning approaches. 

One of the common techniques found in the literature review is state estimation, where information obtained from data sources is used to estimate a state. 
For instance, Viswanath et al.~\cite{viswanath2014design} performed an estimation on the people count of multiple proximity sensors. 
This is also shown in~\cite{Choi2018Bi}, where the infrared and ultra-wideband is used to estimate people count in an urban area.
Another example of the state estimation can be found in~\cite{Weppner2014Participatory}, where it estimates the number of the person using Bluetooth probing sensors in 12 event spaces around the city. 
These types of data extraction techniques mostly focus on the sensor level, and also a singular space information extraction.

To generate space-centric insights about a particular PoI, machine learning approaches have been adopted by many researchers, and it can be divided into supervised and unsupervised machine learning.
Supervised machine learning often includes data with labels or ground-truth, and the most common problems addressed are prediction and classification.
Examples of the prediction approach for space-centric systems can be found in~\cite{tyndall2016occupancy} and~\cite{candanedo2016accurate}. 
Meanwhile, classification techniques have been shown in the following works~\cite{barchiesi2015acoustic,akbar2015contextual,Kleiminger2015Household}, which use classifiers such as neural network and regression.
On the other hand, unsupervised machine learning approaches do not involve data with labels or ground-truth, and common techniques used are clustering as well as dimension reduction.
For instance, the space-centric space monitoring involves clustering techniques, which can be found in~\cite{salamon2015unsupervised} as well as~\cite{FernandezVilas2019Analysis}, where cluster techniques such as $k$-means and DBSCAN are utilized.

After going through the techniques, we observe that the approach mostly are data-driven. 
For instance, if acoustic and vision sensors frequently have ground truth for validation, supervised machine learning would be a more appropriate method for such tasks instead. 
\vspace{-0.23cm}

\begin{figure}[h]
	\includegraphics[width=0.435\textwidth]{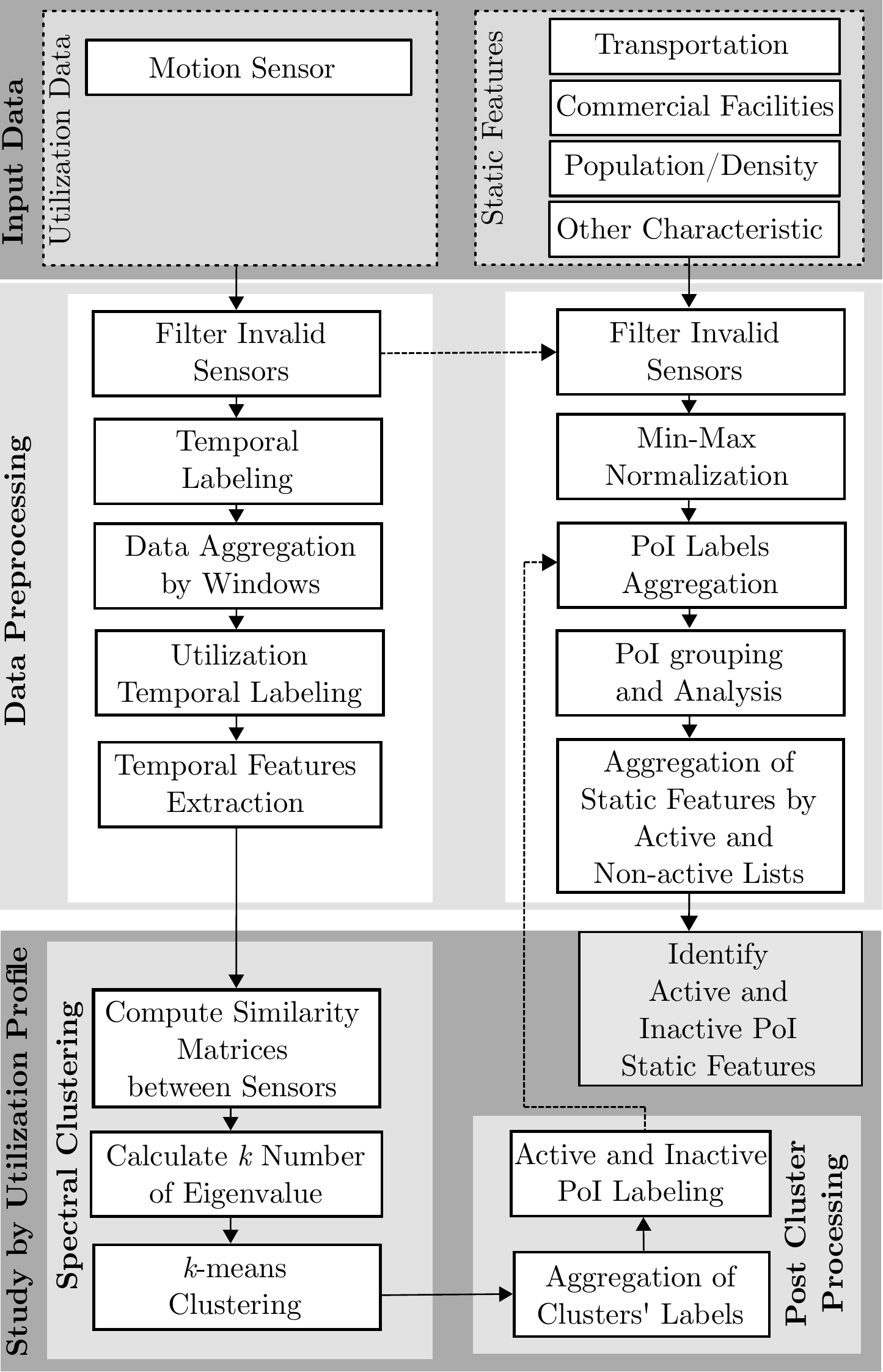}
	\centering \caption{The system model for extracting insights for public space, which consists of three stages: input data, data preprocessing, and sensor profiling}
	\label{fig:data_process}
	\vspace{-0.5cm}
\end{figure}

%%%%%%%%%%%%%%%%%%%%%%%%%%%%%%%%%%%%%%%%%%%%%%%%%%%%%%%%%%%%%%%%%%%%%%%%%%%%%%%%%%%%%%%%%%%%%%%%%%%%%%%%%%
\section{Data Processing Pipeline}
\label{sec:dataPipeline}
%%%%%%%%%%%%%%%%%%%%%%%%%%%%%%%%%%%%%%%%%%%%%%%%%%%%%%%%%%%%%%%%%%%%%%%%%%%%%%%%%%%%%%%%%%%%%%%%%%%%%%%%%%
In this section, we describe the overall data process system, data preprocessing stage, and the modules used for extracting insights from the data.
The system model is used for profiling different public spaces based on their utilization comprises three processing phases, which are (1) data sources, (2) data preprocessing, and (3) utilization pattern clustering and analysis.
The overall system architecture is illustrated in the following Fig.~\ref{fig:data_process}.

\subsection{Data Sources and Characteristics}
There are two data sources needed in the data processing pipeline, which are public space utilization data and static features.
The public space utilization data is collected via sensor nodes proposed in~\cite{lau2018sensor}, while the static features are calculated based on the geographical features of the region studied.

We describe the details about the sensor nodes, whose function is to collect the environmental and utilization data.
There are a total of $47$ valid sensor nodes within the time frame ranging from $01$ May $2017$ to $30$ December $2017$.
The sampling rate for the sensor nodes is tuned to $5$-minute to preserve the longevity of the sensor nodes during data collection.
A minimum requirement $10$\% validity data(based on the utilization data) of the designated time frame is chosen to identify the invalid sensor nodes.
This step will ensure that those sensors with insufficient data entries will be excluded from the data processing pipeline.
Similar invalid sensor lists will be applied to the static features to remove invalid static features.

The static features consist of quantitative data that describes the geographical data for each PoI, where the same location of sensor nodes are deployed, which can be generally divided into the following segments: (1) transportation, (2) commercial facilities, (3) population density, plus (4) other characteristics.
Note that the static features do not have a temporal effect unless there is a modification to the amenities, where it remains constant during the data collection period. 
Detailed information about the static features will be further discussed in Section \ref{sec:magnitudeUtz}.

\subsection{Data Preprocessing}
Two types of input data need to be preprocessed, which are motion sensors and static features. 
The motion sensors will undergo a few preprocessing steps before feeding it to the sensor profiling module.
We apply the temporal label to the data based on the day of the week, which are Monday, Tuesday, Wednesday, Thursday, Friday, Saturday, and Sunday.
Special occasions such as school holidays and public holidays are also included as part of the temporal labeling. 
After that, data aggregation by windows (data over a day) is performed to obtain the average utilization data for each temporal label. 
Next, the public space utilization data through generalized temporal labels (weekday, weekend, and a school holiday) is averaged in order to obtain an individual profile for each generalized temporal label.

As for static features, the invalid sensors list is applied to the static features based on the identification process from motion sensors data processing.
We perform the min-max normalization function using the following equation:
\begin{equation}
Z'=\frac{z_i-min(Z)}{max(Z)-min(Z)}
\label{eqn:minmax}
\end{equation}
where $Z$ is the list of active or non-active sensor nodes and $z_i$ is the static feature of the sensor node being computed.
After that, the inverse function is applied to calculate the distance of normalized static features such as:
\begin{equation}
Z' = 1 - Z'_{d}
\end{equation}
where $Z'_{d} $ is the normalized distance features.
Note that lower distance value denotes higher values in the normalized range, while higher value implies the opposite (notation of $1.0$ shows the nearest distance, where $0.0$ indicates furthest).
After that, we label the features based on the activeness label obtained from the public space utilization profiling module. 
Labeled features will be used for inference analysis in the analysis section.

\subsection{Sensor Profiling and Analysis}
After public space utilization is preprocessed, we profile the sensors based on their utilization to determine their activeness.

The activeness for each PoI based on their public space utilization is studied by using spectral clustering~\cite{Fortunato2010Community} to form different usage profiles of the PoI.
These different profiles of PoI's utilization are three generalized temporal labels, which are weekday, weekend, and school holiday.
The underlying principles of Spectral Clustering will be outlined in the upcoming section.
After obtaining the clustered result, the clusters are separated into an active and non-active list based on the clusters for each generalized temporal label.
The method used to separate the clusters rely on the average normalized public space utilization from each cluster.

Based on the active and non-active list of PoIs, the static features are studied and correlated with the normalized utilization of each PoI to perform a heuristic study. 
The static features can be divided into four main categories, which are (1) transportation, (2) commercial facilities, (3) population density, plus (4) other characteristics.
A heuristic approach is proposed to study the correlation between static features and the activeness of a PoI.

%%%%%%%%%%%%%%%%%%%%%%%%%%%%%%%%%%%%%%%%%%%%%%%%%%%%%%%%%%%%%%%%%%%%%%%%%%%%%%%%%%%%%%%%%%%%%%%%%%%%%%%%%%
\section{Sensor Profile Clustering}
\label{sec:sensorProfiling}

In this section, we discuss the profiling module that groups the sensor nodes based on the normalized utilization values.
The profiling module performs spectral clustering~\cite{von2007tutorial} on the sensor modules by utilizing the concept of a fully connected graph, where the similarity metrics represent the edge of a graph.
The main advantage of such an approach is to allow the increment of sensor nodes in future expansion, and recalculation of the similarity metrics can be done in a straight forward manner.

\subsection{Public Space Similarity Study}
First, the public space normalized utilization data is defined as $\textbf{X} \in \mathcal{R}^{n \times t}$ input matrix as follows: 
\begin{equation}
\textbf{X} = \left[ {\begin{array}{*{20}{c}}
	{{x_{1,1}}}&{{x_{1,2}}}& \ldots &{{x_{1,t}}}\\
	{{x_{2,1}}}&{{x_{2,2}}}& \ldots &{{x_{2,t}}}\\
	\vdots & \vdots & \ddots & \vdots \\
	{{x_{n,1}}}&{{x_{n,2}}}& \cdots &{{x_{n,t}}}\\
	\end{array}} \right],
\end{equation}
where matrix $X$ is the data representing the number of sensor nodes with varying spatial characteristic, $n$, and temporal characteristic is denoted by $t$. 

Subsequently, the raw data with the day of the week is labeled to indicate the different normalized utilization values across the week.
In addition, special occasions such as public holidays and school holidays are labeled as well.
There are nine temporal labels can be generated, which are (1) Monday, (2) Tuesday, (3) Wednesday, (4) Thursday, (5) Friday, (6) Saturday, (7) Sunday, (8) School Holiday, along with (9) Public Holiday.

Next, the normalized utilization data is aggregated through an average function for individual temporal labels.
The average function fits the normalized utilization data into daily average windows consists of $288$ samples (sampling frequency of $5$ minutes) for the aforementioned nine temporal labels.
The average function is defined as follows:
\begin{equation}
\label{eqn:windowsAveraging}
\textbf{X}' = \sum_{i=1}^{n} \frac{1}{d} \left[\sum_{j=1}^{t} x_{i,j:j+d}\right]_{1}^{9}
\end{equation}
where $d$ represents the total sampling data in a particular day of the nine temporal labels.
The main reason for averaging the normalized utilization value over nine temporal labels is to obtain a generic pattern of the normalized utilization for different types of days.
This step is crucial to provide some metrics for each sensor to indicate their regular normalized utilization pattern for us to further investigate. 

Based on the individual regular normalized utilization values, the similarity kernels can be formulated to inspect whether both sensor nodes yield similar utilization patterns.
Here, Windows Inverse Euclidean Distance (WIED) function, \textit{dist}$(x'_{a}, x'_{b})$ is applied to address non-linearity of the data as follows: 

\begin{equation}
\label{eqn:SimilarityMatrix}
\text{\textit{dist}}(x_{a}, x_{b}) = \frac{1}{t}\sum_{j=1}^{t} \sqrt{(x'_{a,j} - x'_{b,j})^2},
\end{equation}
where $a$ and $b$ represent the any pairwise of the windowed sensor $(a,b) \in \{1,2,...,n\}$ in $x'$, while $a\ne b$. 
The time windows, $W$ consider previous and subsequent the normalized utilization data to calculate the similarity measurement for both comparing sensor node values within the time-frame of $W$.
By iterating the WIED function over all pairwise sensor nodes, the pairwise similarity   $\textbf{S}_{\textbf{f}q}$ can be calculated by inverting the distance function as such:
\begin{equation}
\label{eqn:inverseDistance}
\textbf{S}_{\textbf{f}q} =  \frac{1}{1+\textit{dist}(x_{a}, x_{b})},
\end{equation}
where $q$ denotes the number of temporal features label, while $\textbf{f}$ is the temporal feature.
Note that if both sensor nodes' utilization data are identical, the similarity measurement will be closer to $1$, where any difference between sensor nodes will be exponentially increased.

By combining the similarity measurement for pairwise sensor nodes, we are able to obtain the similarity matrix $\textbf{S}_{\textbf{f}q}$ of normalized utilization pattern as follows:
\begin{equation}
\textbf{S}_{\mathtt{U}} = \left[ {\begin{array}{*{20}{c}}
	{0}&{{s_{1,2}}}& \ldots& {{s_{1,n-1}}} &{{s_{1,n}}}\\
	{{s_{2,1}}}&{0}& \ldots& {{s_{2,n-1}}} &{{s_{2,n}}}\\
	\vdots & \vdots & \ddots& \vdots  & \vdots \\
	{{s_{n-1,1}}}&{{s_{n-1,2}}}& \cdots &0 &{{s_{n-1,n}}}\\
	{{s_{n,1}}}&{{s_{n,2}}}& \cdots &{{s_{n,n-1}}} &{0}\\
	\end{array}} \right].
\end{equation}
The similarity matrix, $\textbf{S}_{\mathtt{U}}$ has an orthogonality property since it is a complete graph and it can be used for spectral clustering algorithm (using similar concept from community detection~\cite{Fortunato2010Community}). 
In this paper, the sensor nodes are represented as graph nodes, and similarity metrics between different nodes are computed as edges.

As for the similarity matrix, a different time slot of the day is examined as well as $24$ hours time-frame similarity. 
However, comparing $24$ hours of difference between every sensor node does not seem effective as a common inactive period between sensor nodes mostly yields higher similarity values.
The similarity values generated during the inactive period (midnight) will likely cause noise when comparing one sensor node to another.
So, the underlying similarity metrics rely on the weight assigned to different periods to mitigate the dominance of similarity values of the inactive period.
We further investigate temporal features in conjunction with similarity matrix, $\textbf{S}_{\mathtt{U}}$ to form affinity matrix $\textbf{A}$ by breaking down the time into $4$ sessions using the following Table~\ref{tbl:temporalTime}. 

\begin{table}[h]
	\centering
	\caption{Temporal Features Time}
	\label{tbl:temporalTime}
	\begin{tabular}{@{}c|cccc@{}}
		\toprule
		section & morning, $f_1$ & afternoon, $f_2$ & evening, $f_3$ & night, $f_4$ \\ \midrule
		time start & 06:00 & 11:00 & 14:00 & 18:00 \\
		time end & 10:59 & 13:59 & 17:59 & 23:59 \\ \bottomrule
	\end{tabular}
\end{table}
By defining the above mentioned temporal features, the similarity matrix is computed for different time slots.
Each time slot generate a similarity matrices such as $ \textbf{S}'_{\textbf{f}1}, \textbf{S}'_{\textbf{f}2}, \textbf{S}'_{\textbf{f}3},$ and $\textbf{S}'_{\textbf{f}4}$.
To show the effectiveness of including the temporal features, a toy example that consists of two sensor nodes data to check the similarity calculation between two sensor nodes utilization values is proposed. 
	The sensors node utilization values are illustrated in following Fig.~\ref{fig:sensorNodescomparison}:
\begin{figure}[h]
	\begin{center}
		\includegraphics[width=0.48\textwidth]{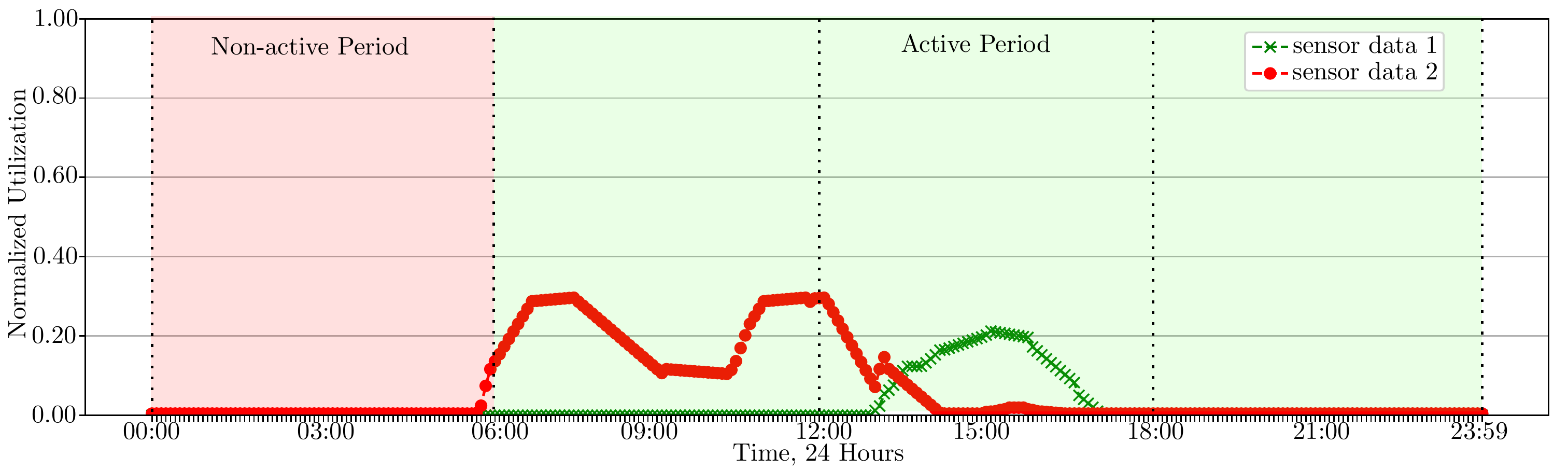} 
		\vspace{-0.2cm}
		\caption{An example of two sensor nodes utilization data, where the red highlight denotes non-active period and the green highlights show active period.}
		\label{fig:sensorNodescomparison}
	\end{center}
\end{figure}

The similarity features with the inclusion of the temporal features or non-temporal features is proposed to show the importance of defining the similarity features based on the toy example above. The result of the comparison is shown in the following Table~\ref{tbl:comparisonSimilarityResult}.
\begin{table}[h]
	\begin{center}
		\caption{Comparison of the similarity calculation using toy example from Fig~\ref{fig:sensorNodescomparison}}
		\label{tbl:comparisonSimilarityResult}
		\begin{tabular}{@{}l|c|c@{}}
			\toprule
			\multicolumn{1}{c|}{\multirow{2}{*}{Similarity Kernel}} & \multicolumn{2}{c}{Pairwise Comparison} \\ \cmidrule(l){2-3} 
			\multicolumn{1}{c|}{} & Without Temporal Features & With Temporal Features \\ \midrule
			Euclidean & 0.9798 & 0.3122 \\
			Manhattan & 0.9167 & 0.0415 \\
			Minkowski & 0.9177 & 0.3121 \\
			WIED & - & 0.7845 \\ \bottomrule
		\end{tabular}
	\end{center}
\end{table}

Based on the observation, we notice that without temporal features, the similarity metrics, the similarity calculation for the majority of the approaches (Euclidean, Manhattan, and Minkowski) are quite high.
However, when the temporal features are introduced, which excluded the weightage of the non-active period, the similarity metrics are greatly reduced.
This shows that the non-active period yields a high proportion of time in computing the similarity between two sensors.
Therefore, we use the WIED as the similarity computation between sensor nodes, which extends the Euclidean features with the windows sliding method.
Note that the similarity value is will be exponentially smaller if the difference between two sensor nodes is large.
After deciding the similarity kernel calculation, the similarity matrix generation used for clustering is discussed in the next sub-section.

\begin{figure*}[ht!]
	\centering
	\begin{subfigure}{1.0\textwidth}
		\begin{tabular}{@{}ccc@{}}
			Weekday & Weekend & School Holiday \\ 
			\includegraphics[width=0.28\textwidth]{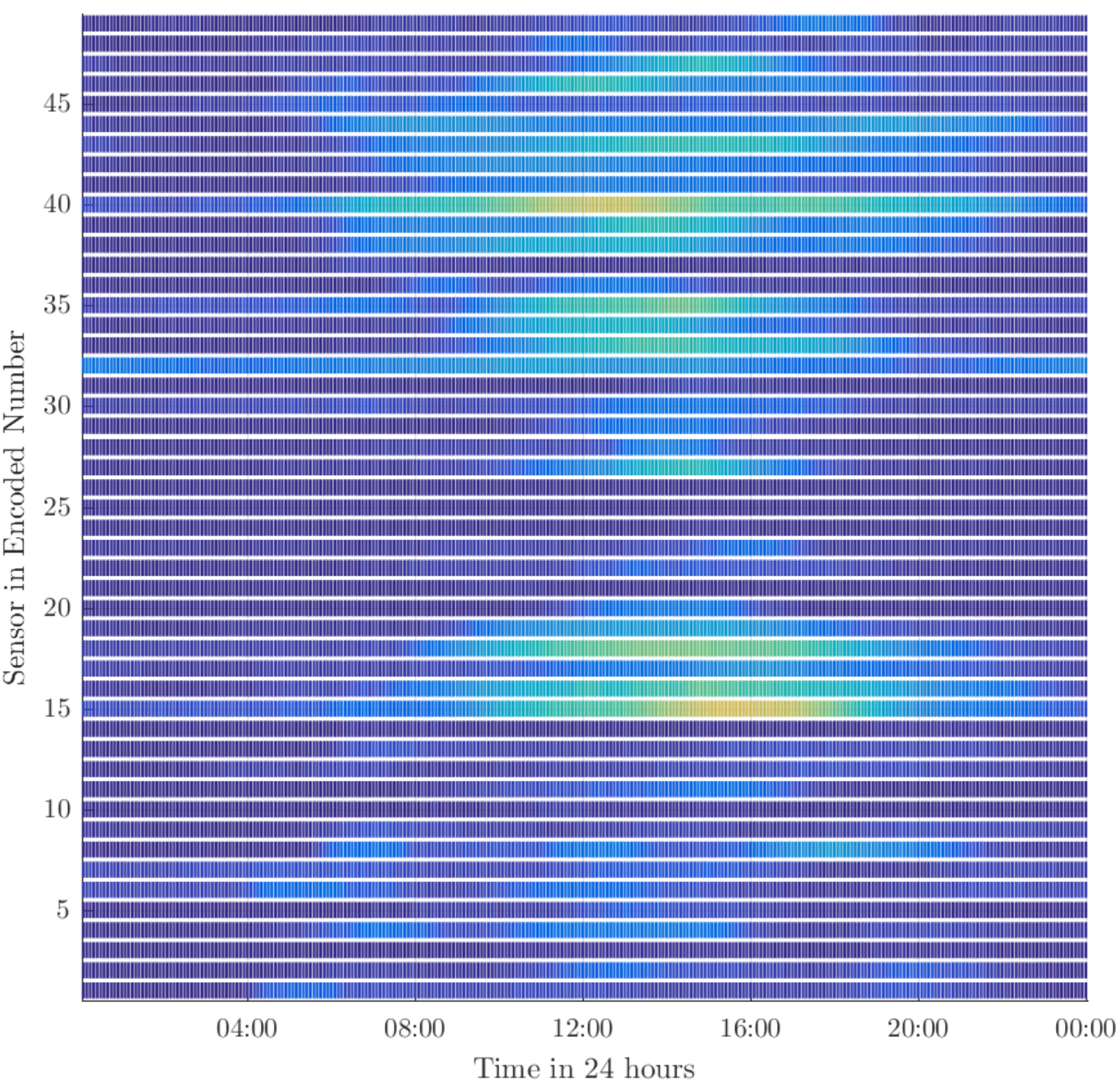}&
			\includegraphics[width=0.28\textwidth]{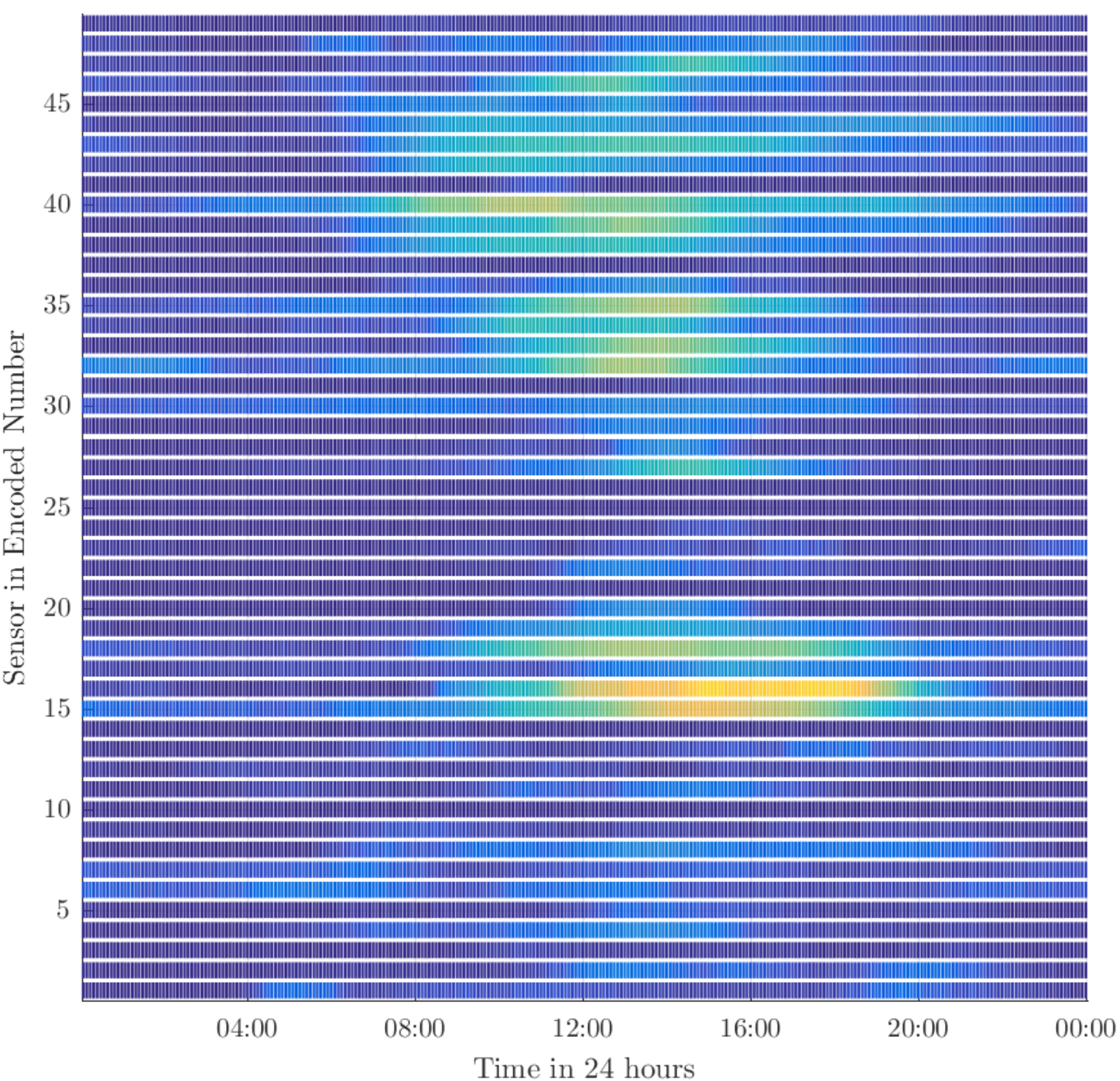}& 
			\includegraphics[width=0.28\textwidth]{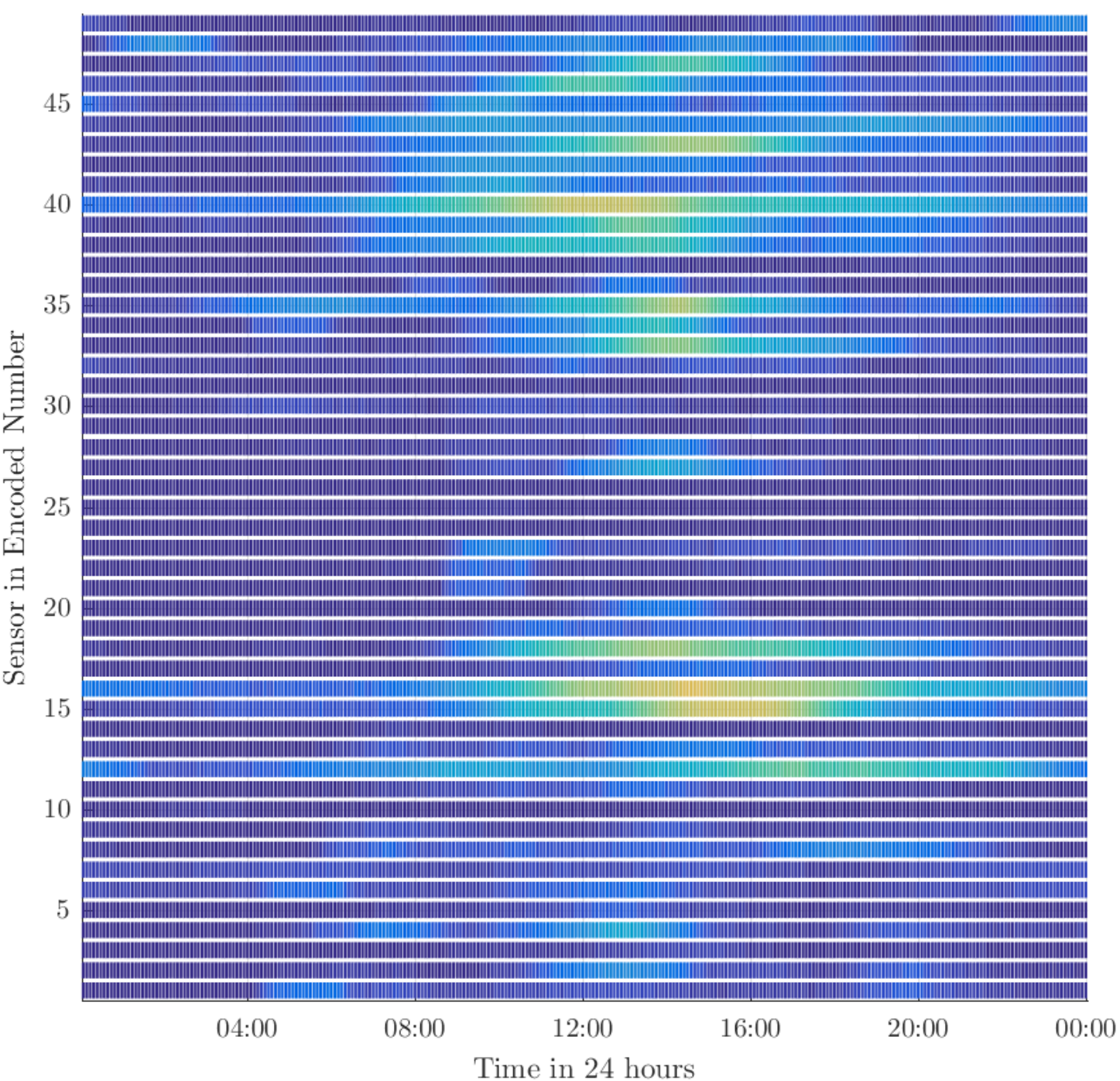} \\ 
		\end{tabular}
		\label{fig:AverageUtz1}
		\centering \caption{Average of public space utilization for three generic temporal labels} 
	\end{subfigure}
	\begin{subfigure}{1.0\textwidth}
		\begin{tabular}{@{}ccc@{}}
			Weekday & Weekend & School Holiday \\ 
			\includegraphics[width=0.28\textwidth]{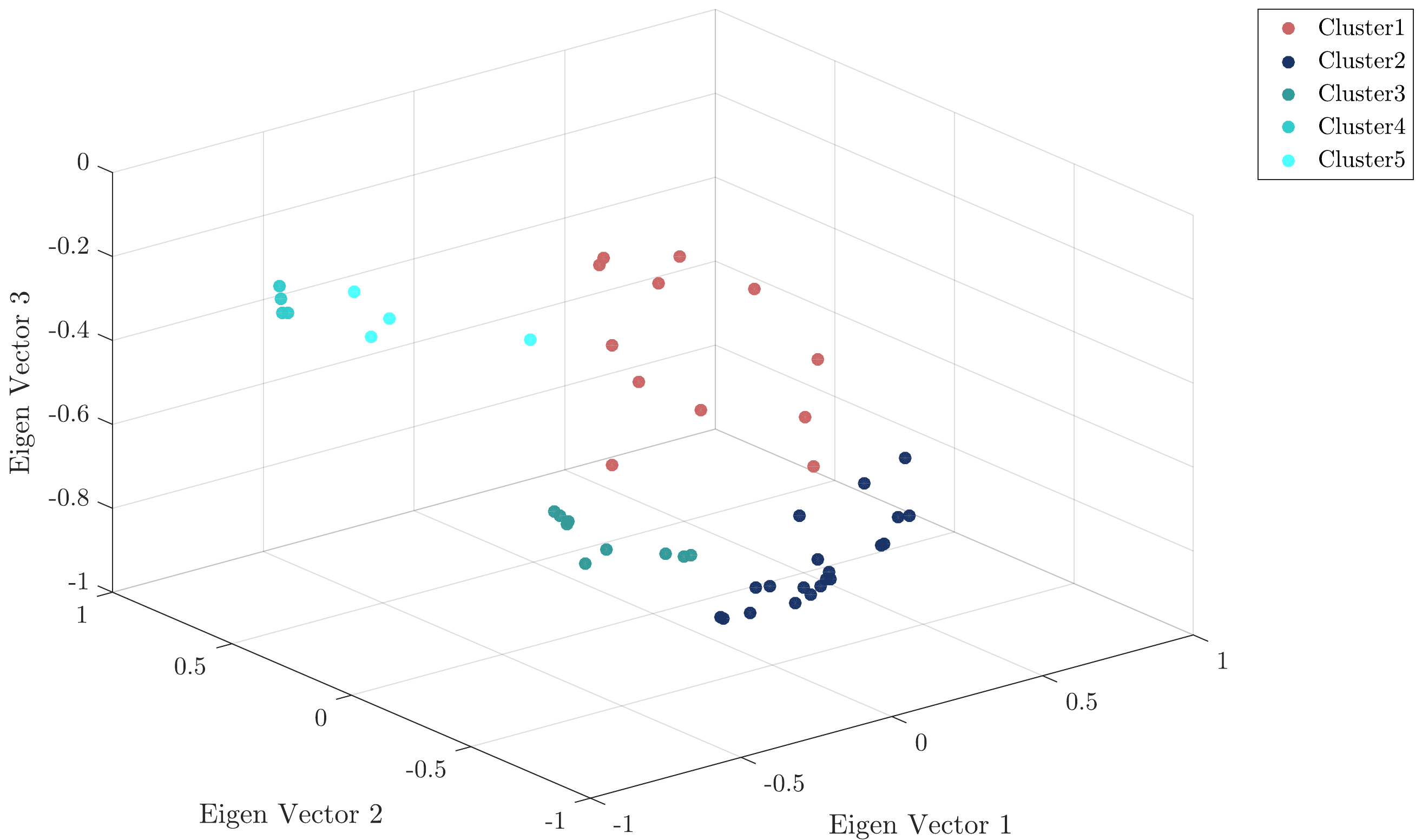}&
			\includegraphics[width=0.28\textwidth]{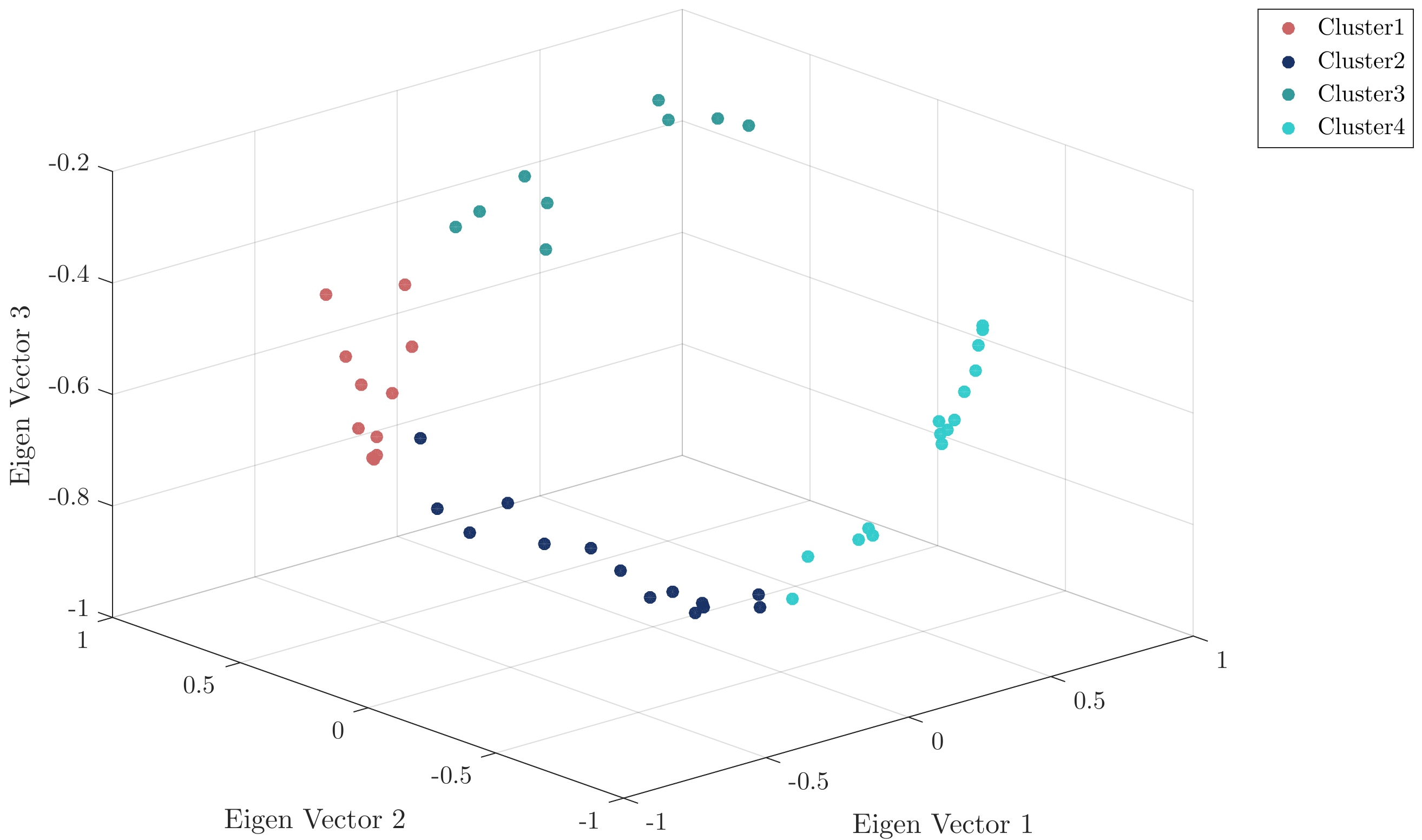}& 
			\includegraphics[width=0.28\textwidth]{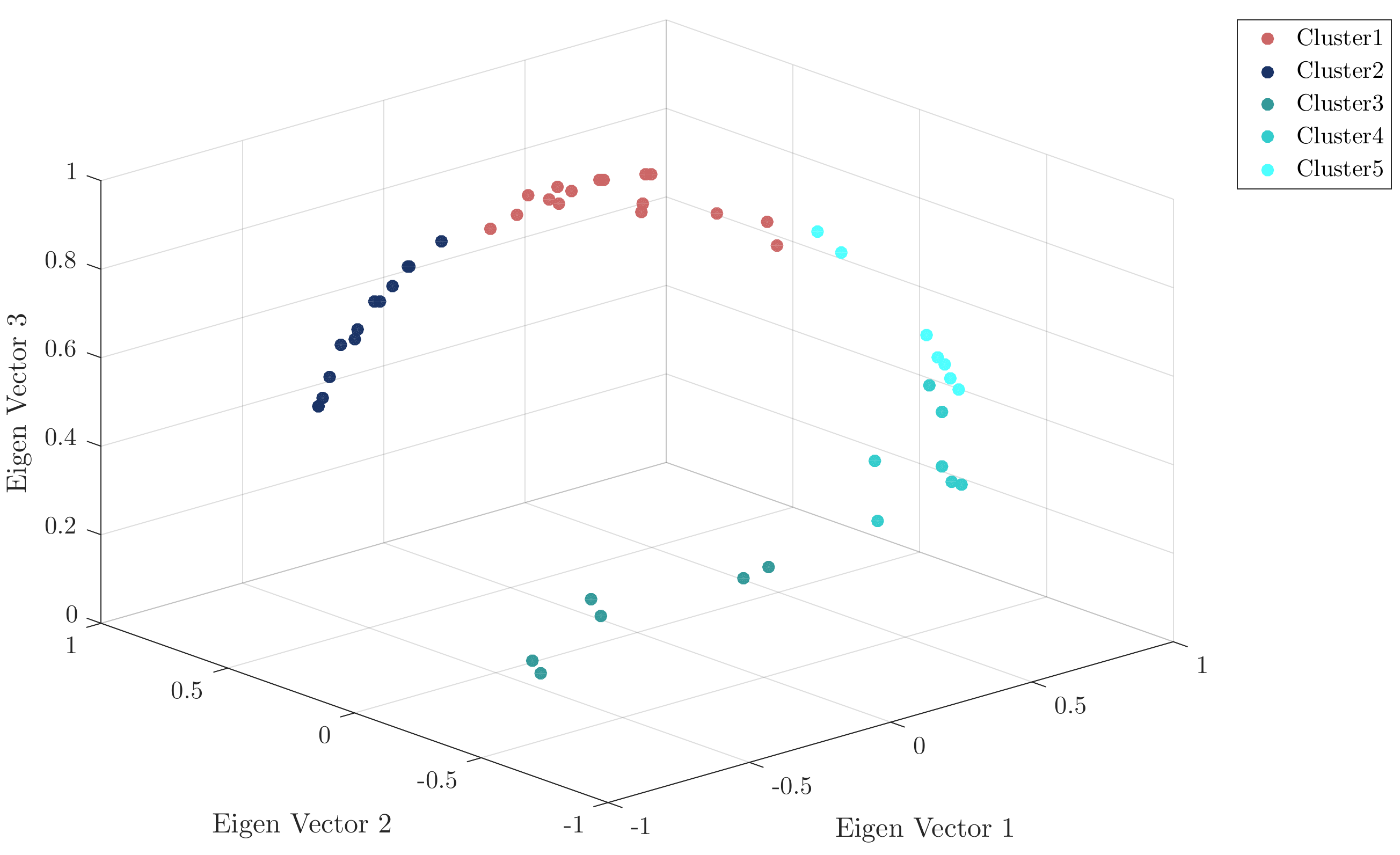} \\ 
		\end{tabular}
		\label{fig:AverageUtz2}
		\centering \caption{Visualization of the Eigen Vector for each generic temporal labels} 
	\end{subfigure}
	\centering \caption{The average data input for each sensor and the Eigen Vectors calculated each generic temporal labels}
	\label{fig:AverageUtz}
	\vspace{-0.4cm}
\end{figure*}

\begin{figure*}[b!]
	\begin{tabular}{@{}c|ccccc@{}}
		\toprule
		& \begin{tabular}[c]{@{}c@{}}Category 1 \\ ($0.30$ and above) \end{tabular}  &
		\begin{tabular}[c]{@{}c@{}}Category 2 \\ ($>0.20$ and $\le0.30$) \end{tabular}&
		\begin{tabular}[c]{@{}c@{}}Category 3 \\ ($>0.10$ and $\le0.20$) \end{tabular} & 
		\begin{tabular}[c]{@{}c@{}}Category 4 \\ ($>0.03$ and $\le0.10$) \end{tabular} & 
		\begin{tabular}[c]{@{}c@{}}Category 5 \\ ($0.03$ and below) \end{tabular} \\ \midrule
		\rot{\hspace{1.1cm}Weekday} &    
		\includegraphics[width=0.17\textwidth]{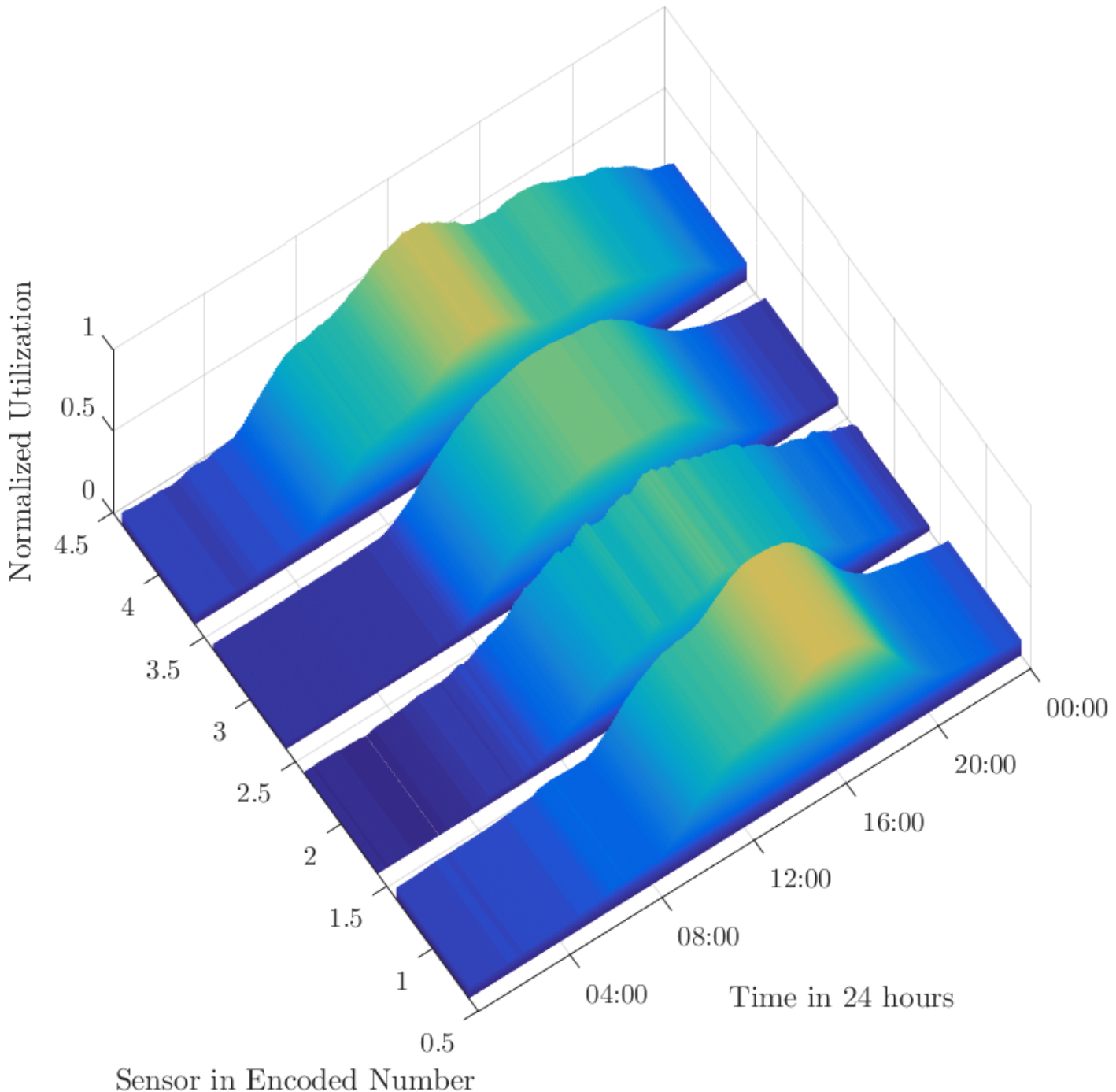}&  
		\includegraphics[width=0.17\textwidth]{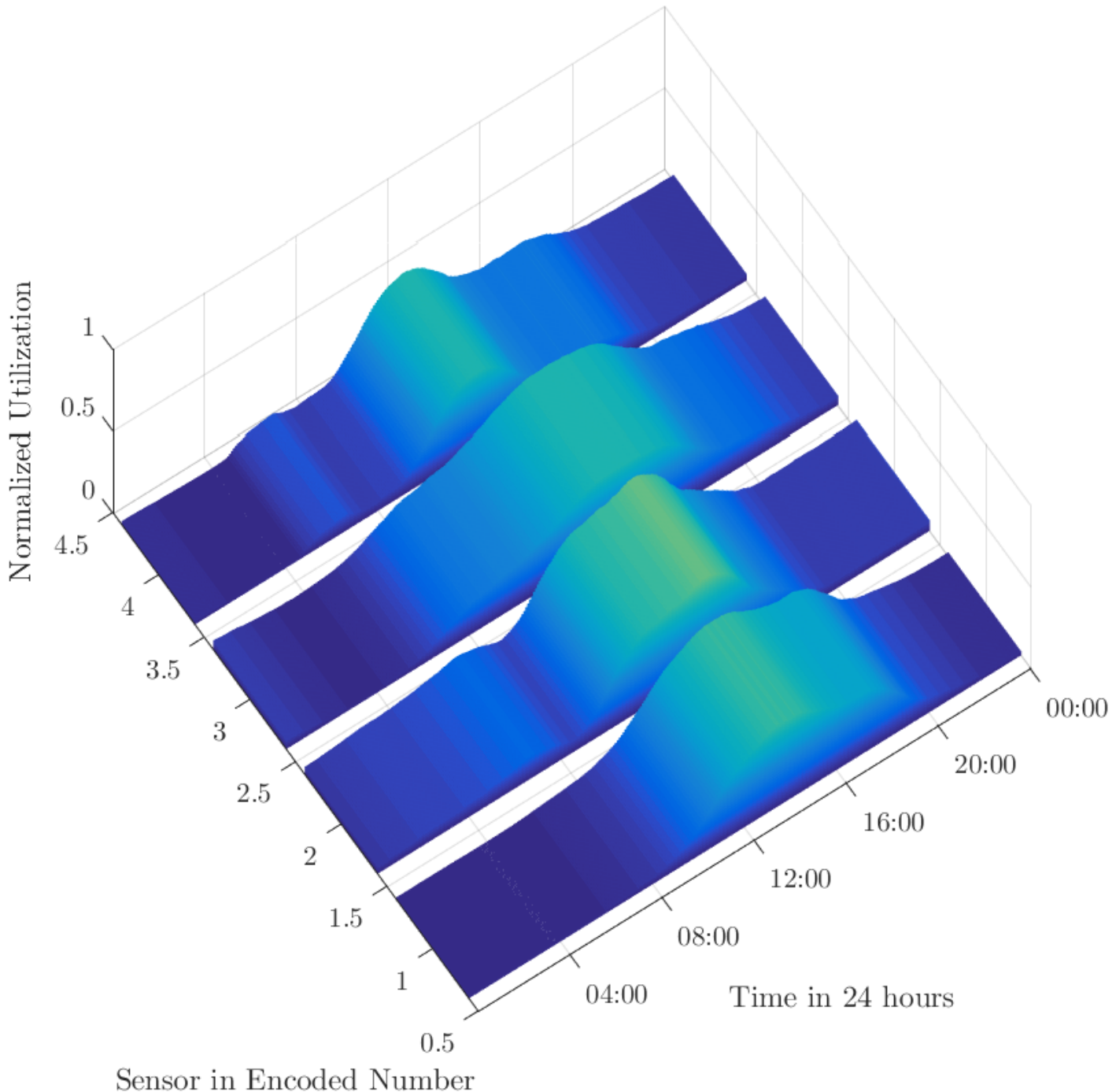}&  
		\includegraphics[width=0.17\textwidth]{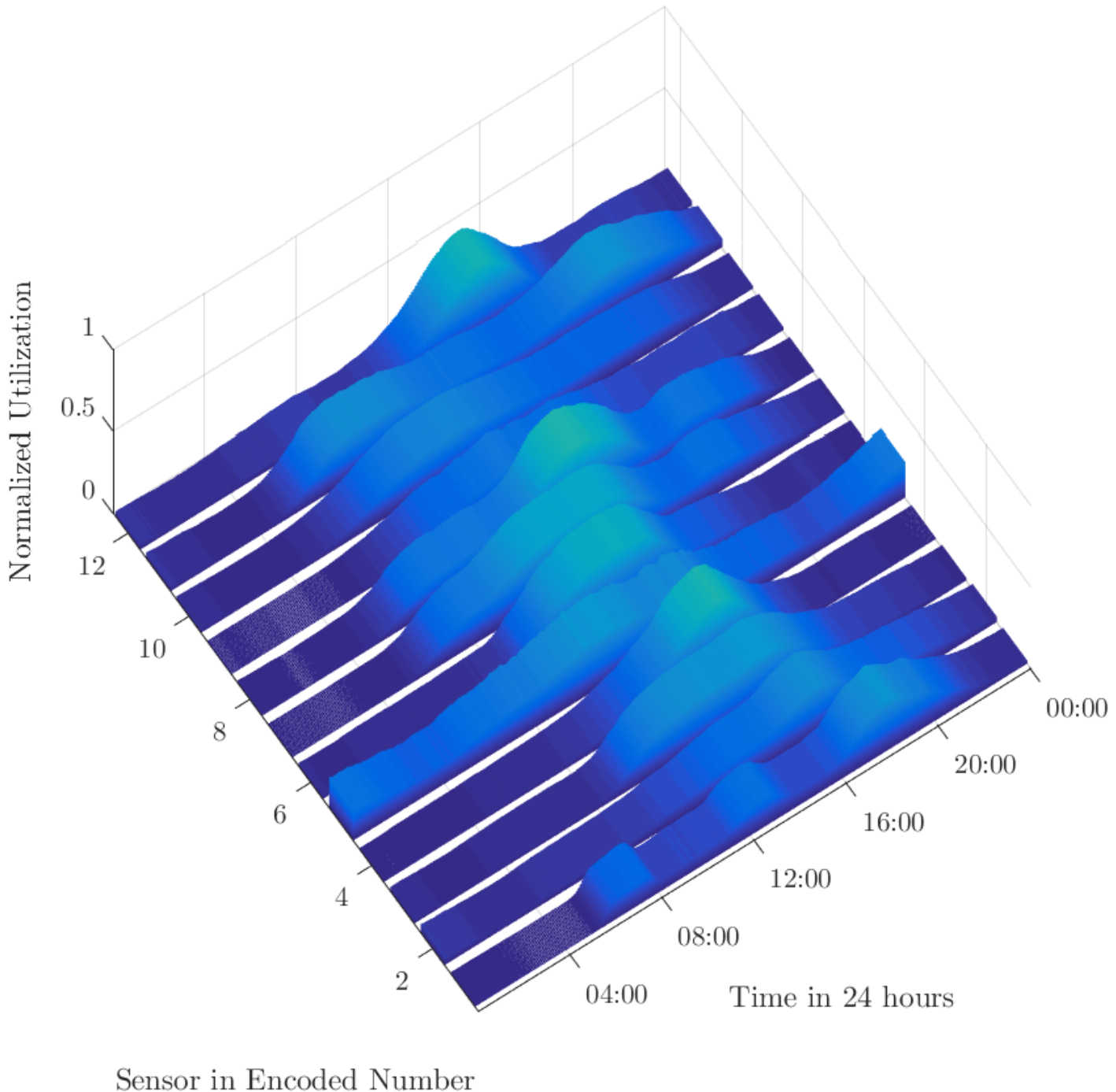}&  
		\includegraphics[width=0.17\textwidth]{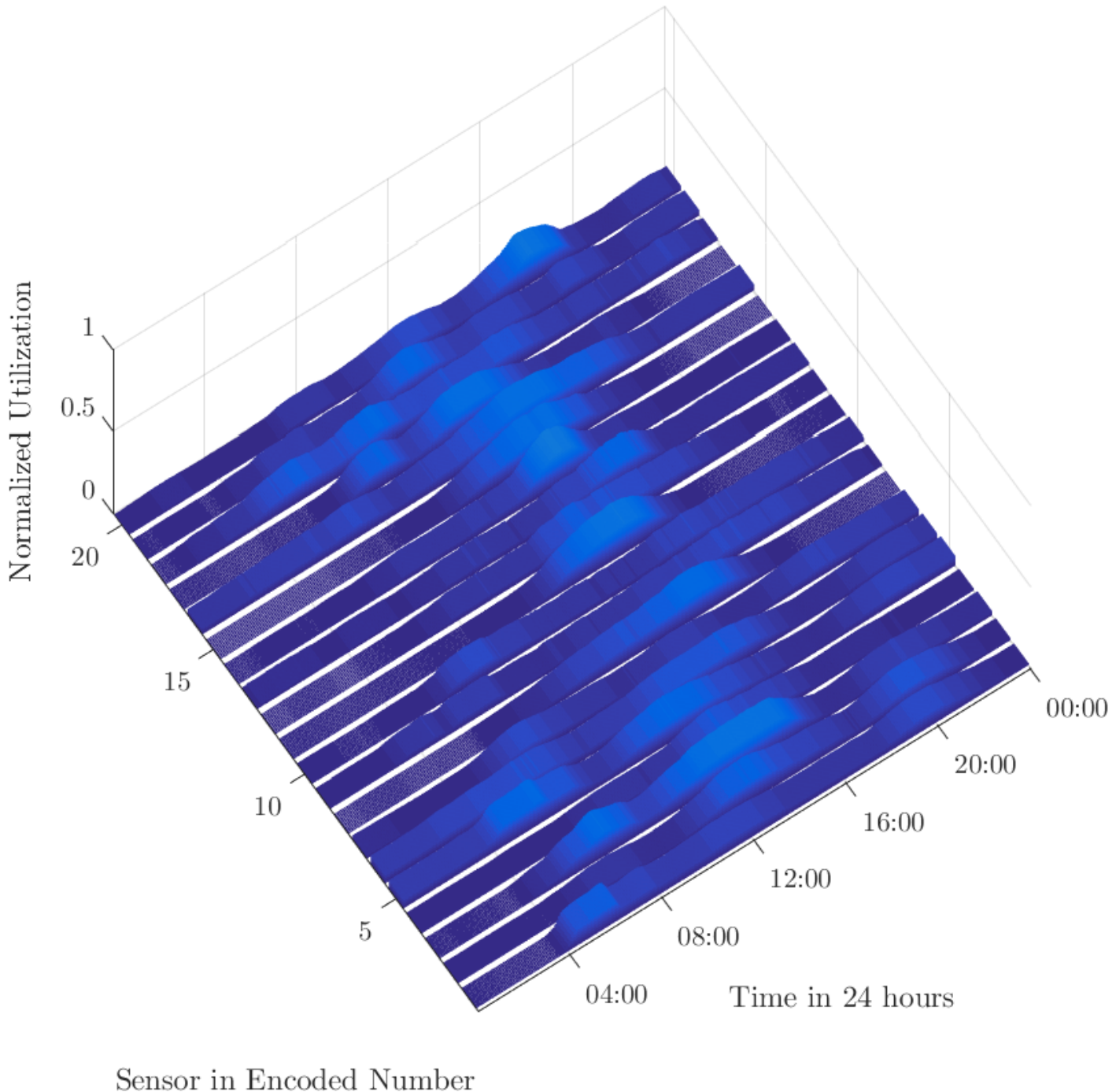}&
		\includegraphics[width=0.17\textwidth]{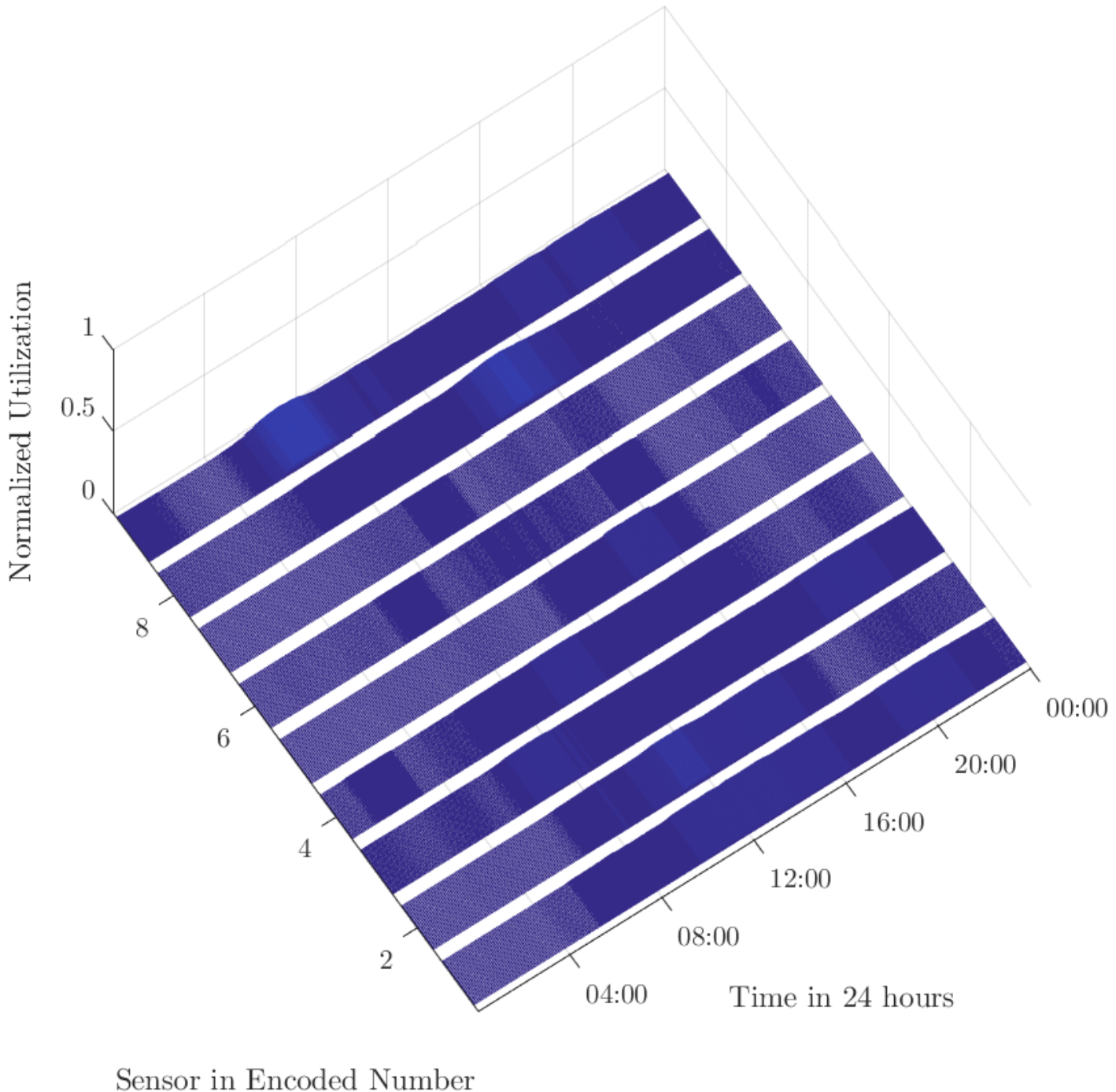}  \\
		&  $\bar{X}=$0.3850 & $\bar{X}=$0.2301 & $\bar{X}=$0.1621 & $\bar{X}=$0.0577 & $\bar{X}=$0.0105 \\ \midrule
		\rot{\hspace{1.1cm}Weekend} &  
		\includegraphics[width=0.17\textwidth]{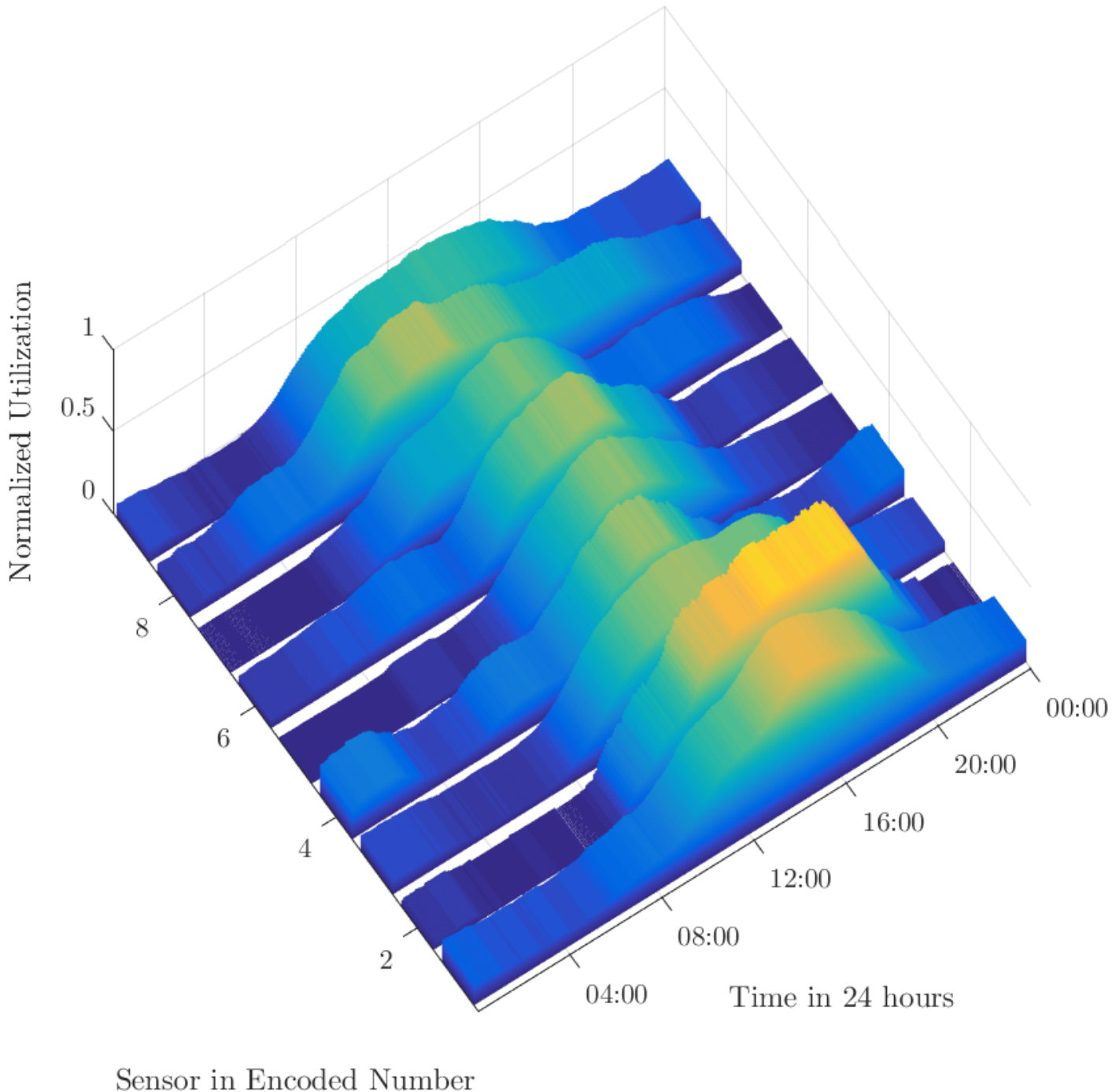}&  
		&  
		\includegraphics[width=0.17\textwidth]{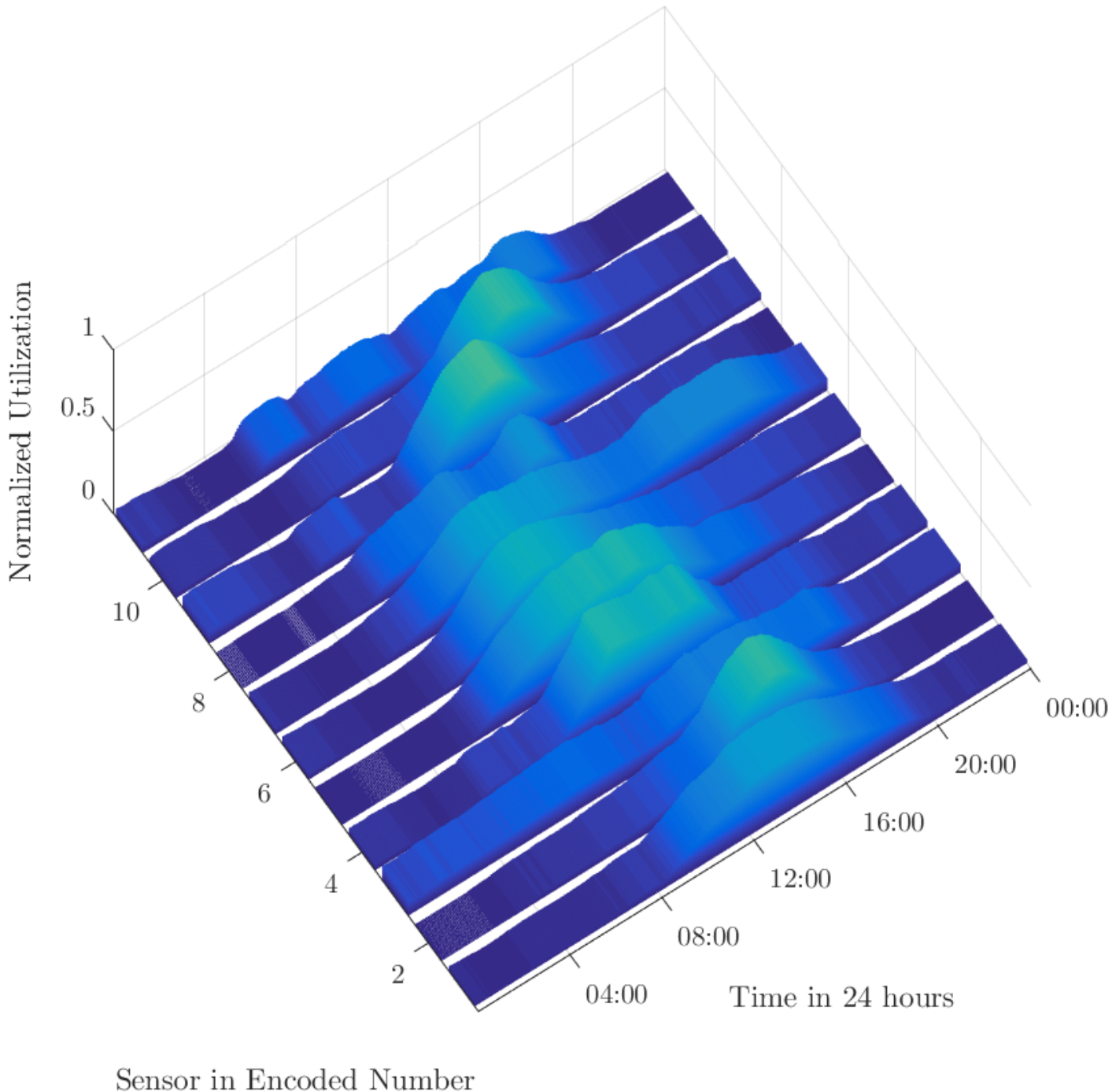}&  
		\includegraphics[width=0.17\textwidth]{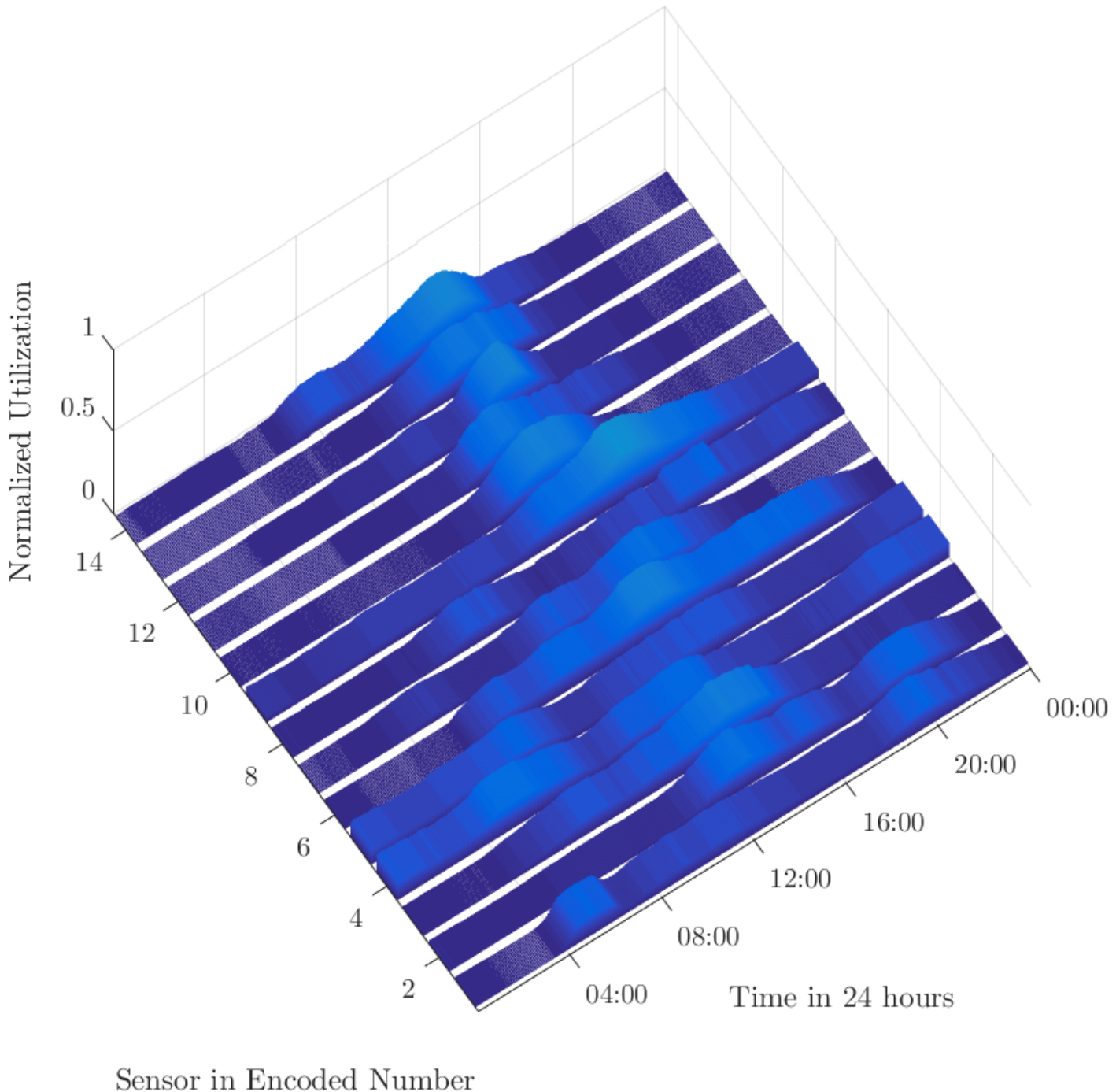}&
		\includegraphics[width=0.17\textwidth]{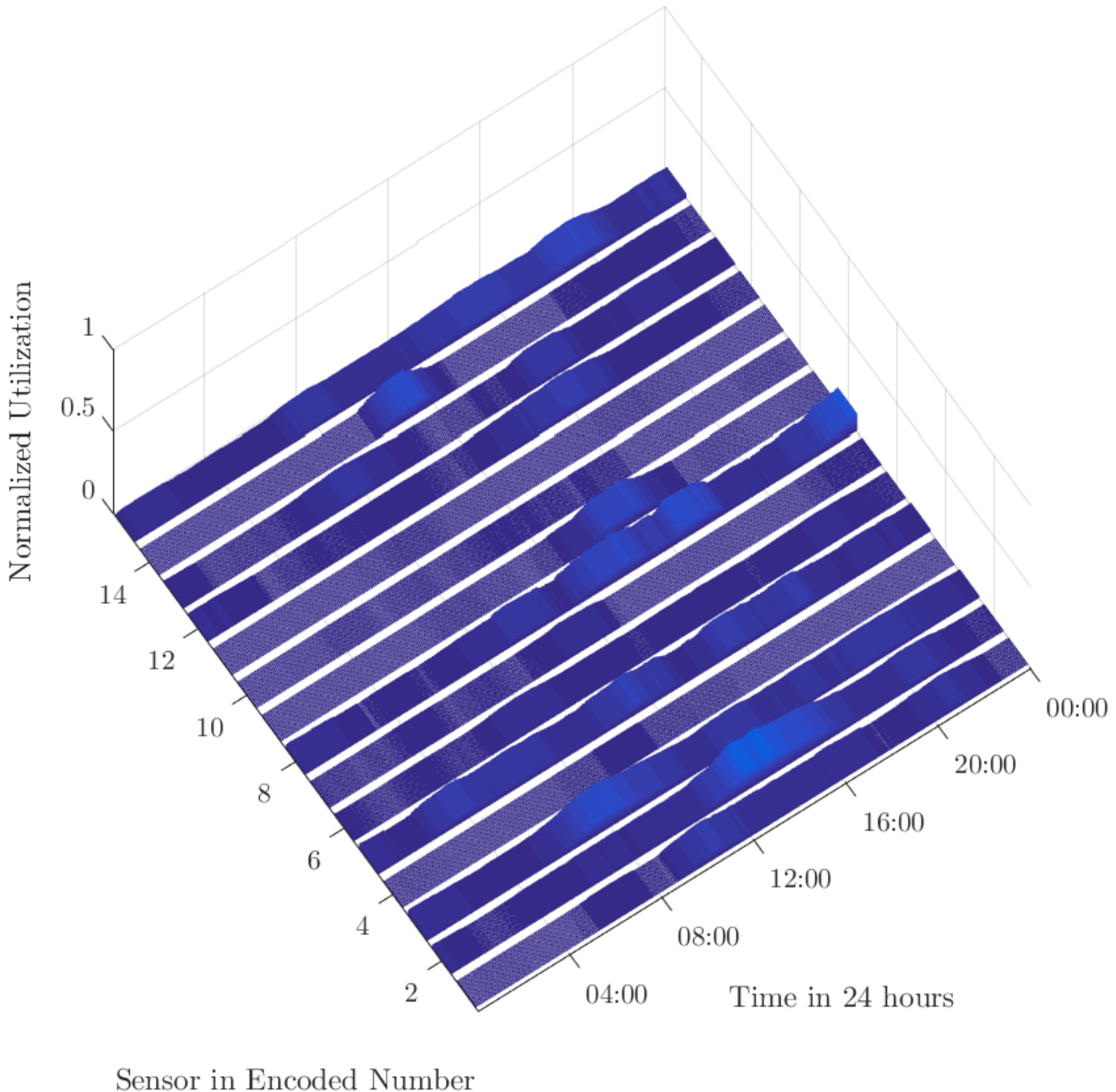}  \\
		&  $\bar{X}=$0.3483 & $\bar{X}=$ - & $\bar{X}=$0.1709 & $\bar{X}=$0.0764 & $\bar{X}=$0.0199 \\ \midrule
		\rot{\hspace{0.2cm}School Holiday} &  
		\includegraphics[width=0.17\textwidth]{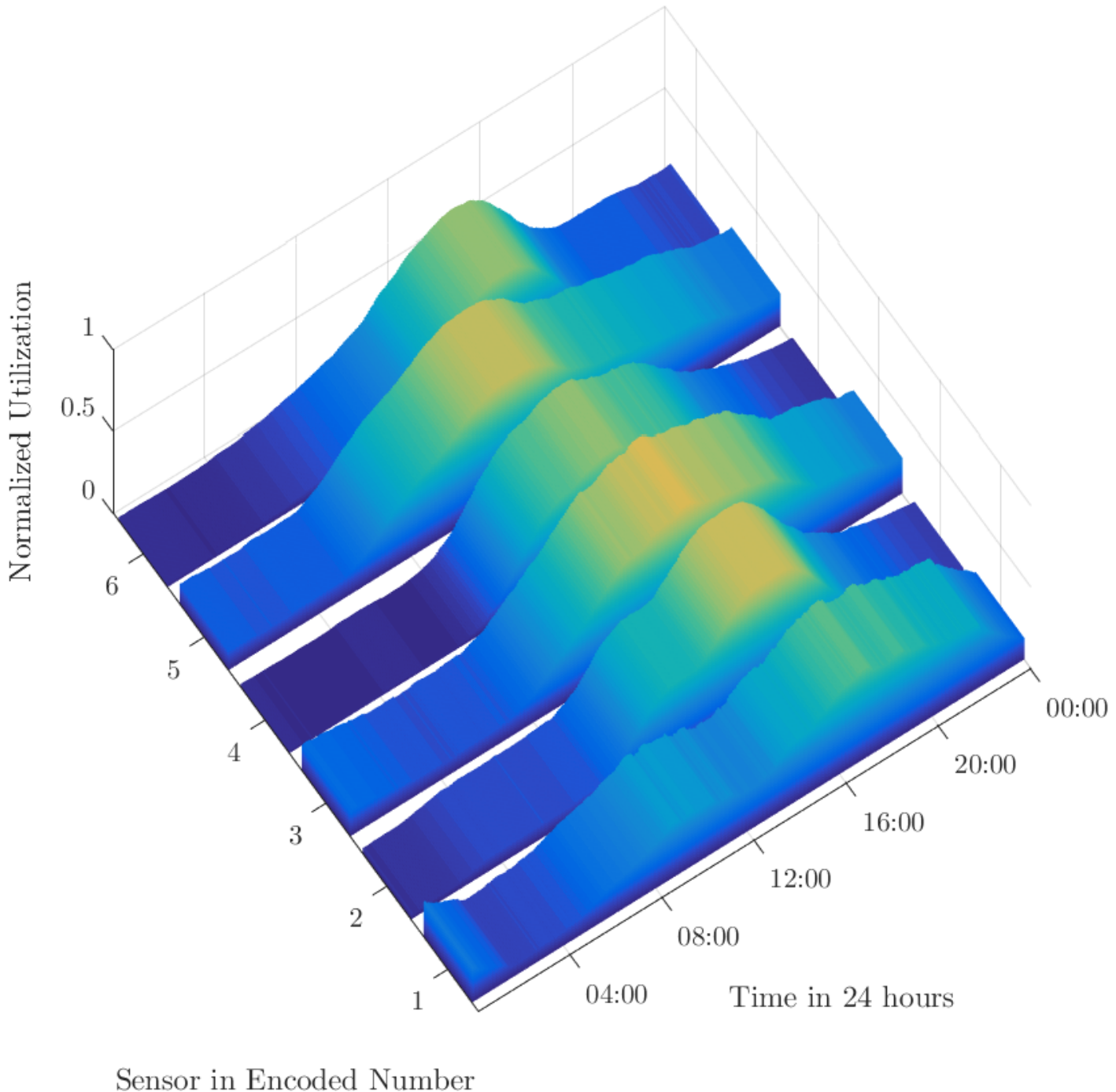}&  
		\includegraphics[width=0.17\textwidth]{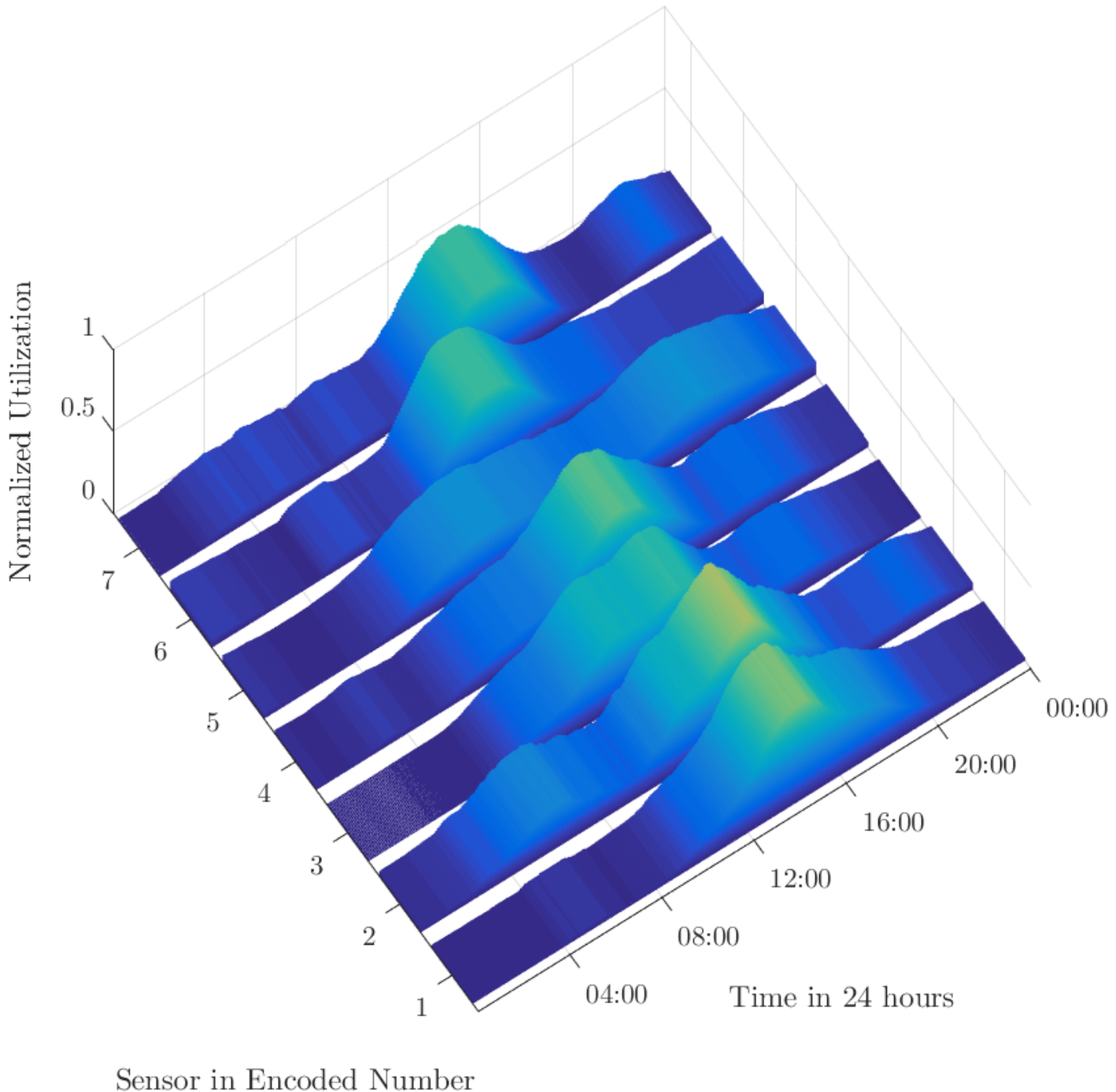}&  
		\includegraphics[width=0.17\textwidth]{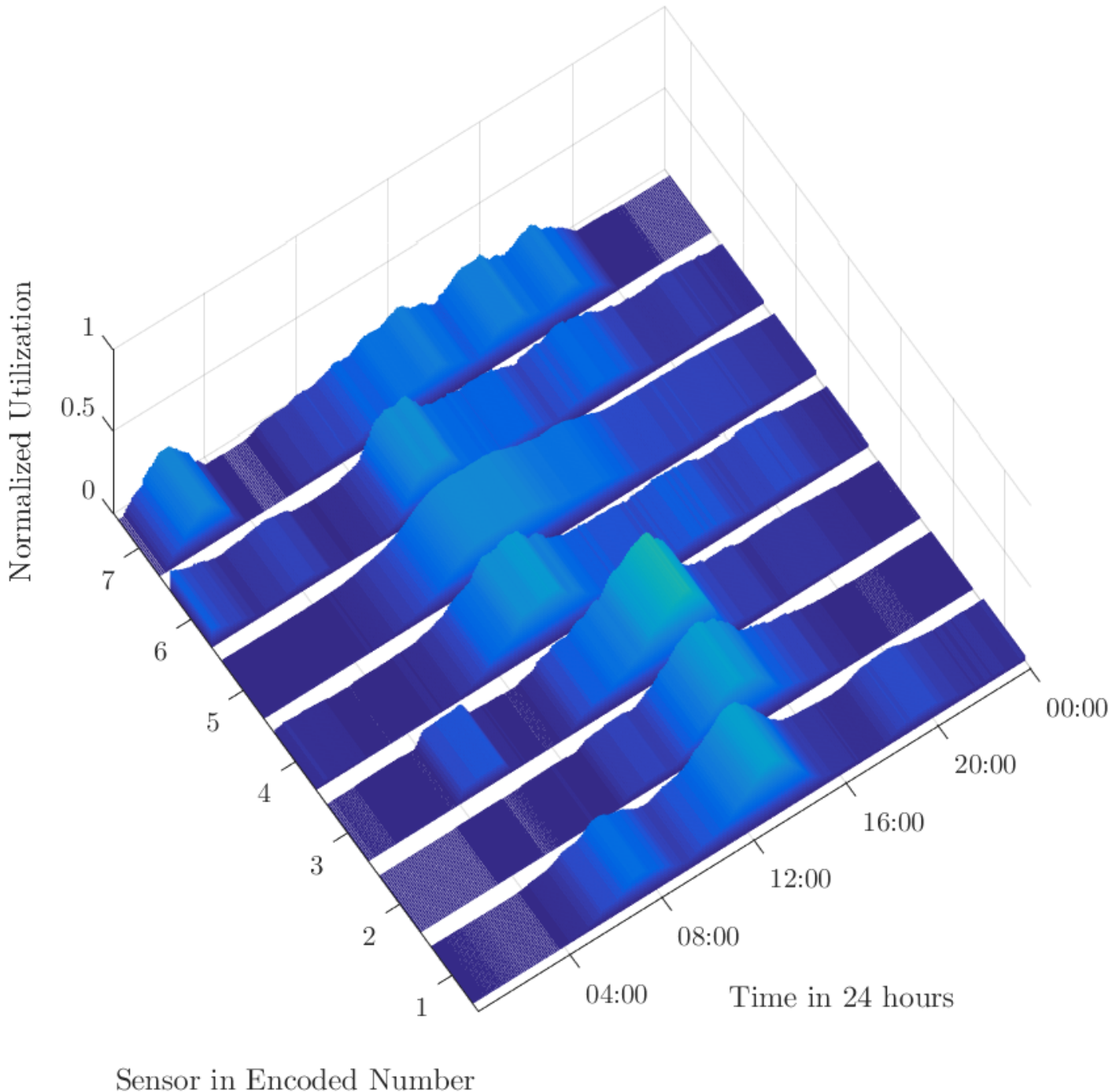}&  
		\includegraphics[width=0.17\textwidth]{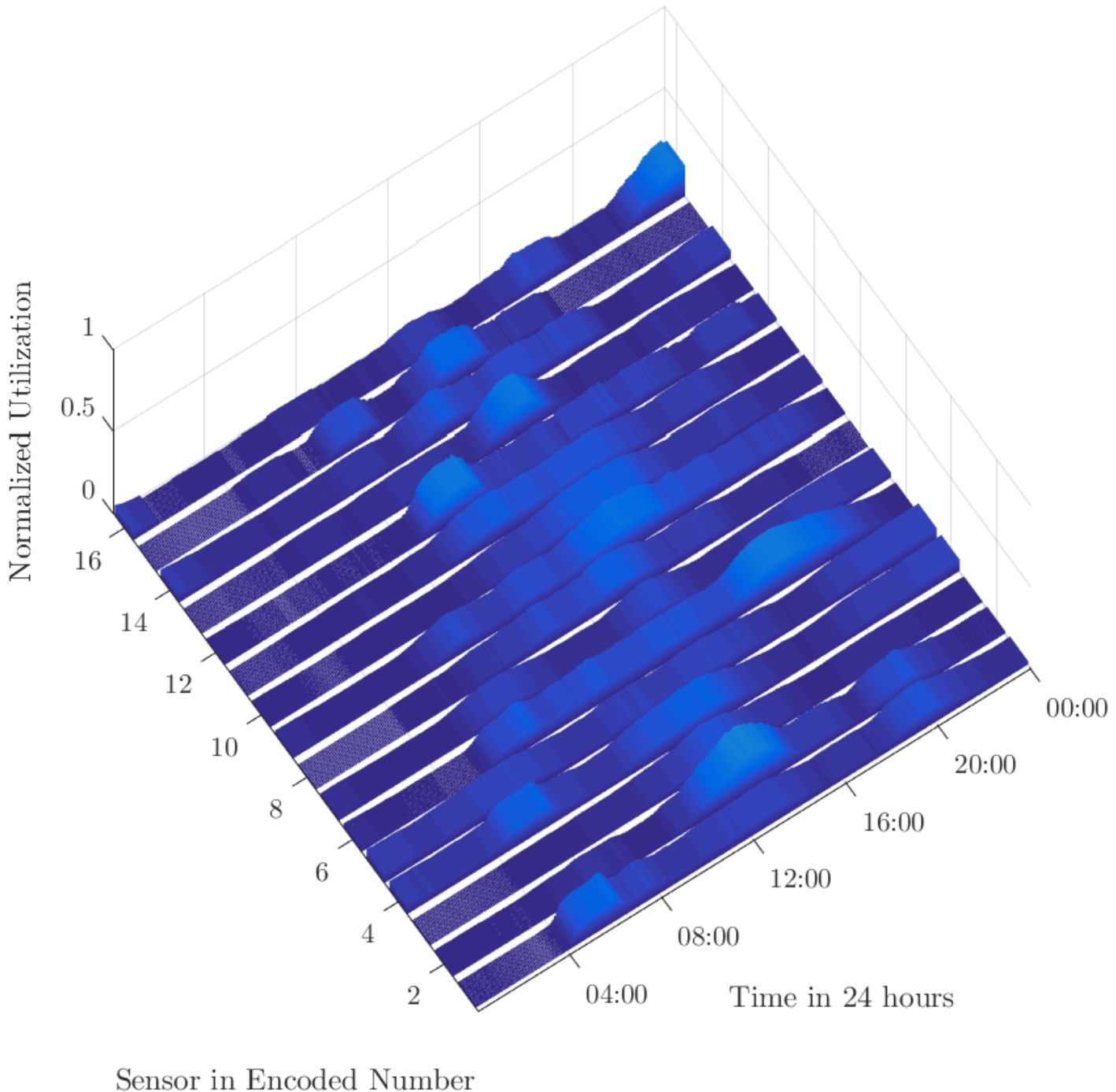}&
		\includegraphics[width=0.17\textwidth]{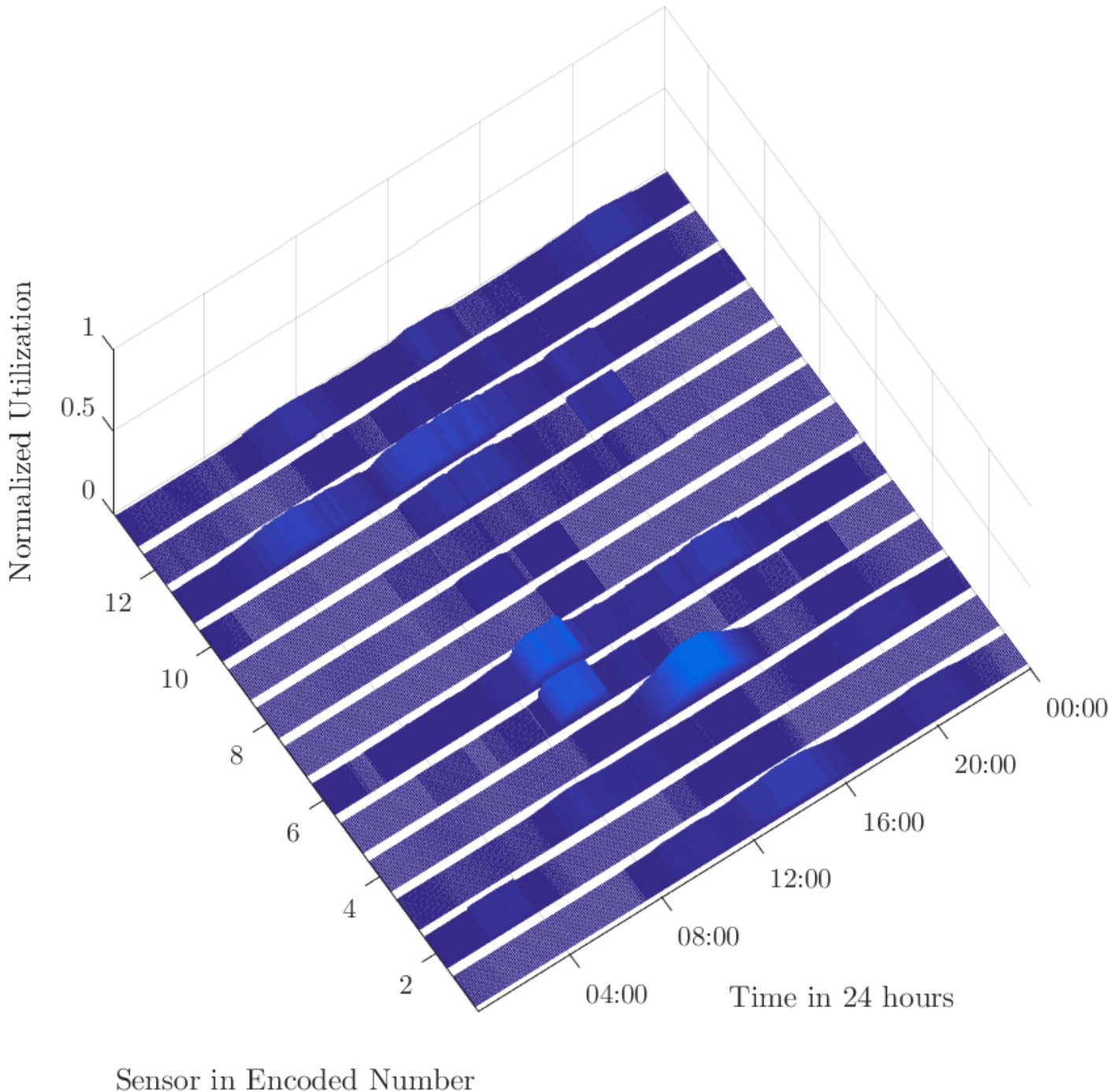}  \\ 
		&  $\bar{X}=$0.3769 & $\bar{X}=$ 0.2138 & $\bar{X}=$0.1139 & $\bar{X}=$0.0578 & $\bar{X}=$0.0150 \\ \bottomrule
	\end{tabular} \\
	%\end{tabular}
	\centering \caption{The clustering results for three different time slots - Weekday, Weekend, and School Holiday.}
	\label{fig:temporalMatrix}
	
	\vspace{-0.4cm}
\end{figure*}

\subsection{Multi Features Spectral Clustering}
In this sub-section, the similarity matrix of temporal features is combined with public space normalized utilization pattern to form affinity matrix, $A$ as follows:
\begin{equation}
\label{eqn:affinityMatrixDerive}
\textbf{A} = \frac{w_1}{4}(\textbf{S}'_{\textbf{f}1} + \textbf{S}'_{\textbf{f}2} + \textbf{S}'_{\textbf{f}3} + \textbf{S}'_{\textbf{f}4}) + w_2 \textbf{S}_{\mathtt{U}}
\end{equation}
where $w_1$ and $w_2$ denote the weights for similarity matrix and temporal features respectively, where $w_1 + w_2 =1.0$.
In the data processing pipeline, uniform weight is considered ($w1 = w2 = 0.5$) to generate the affinity matrix $A$. 

In order to compute the Laplacian matrix, $L$ for generalized eigenvectors, the degree matrix, $D$ matrix can be denoted as follows: 
\begin{equation}
D = \sum_{i}^{n} \left\{ {\begin{array}{*{20}{c}}
	1 & \textbf{A}_{i,i} \ge 0  \\
	0 & \text{ otherwise}
	\end{array}} \right. ,
\end{equation}
where it accounts a non-zero affinity matrix (complete dissimilar matrix) and similarity metrics within the nodes.
It is also commonly denoted as $deg$(), and it calculates the number of connected nodes within a graph.

Finally, by combining degree and affinity matrix, the normalized Laplacian matrix, $\textbf{L}$ can be formulated using the following equation:
\begin{equation}
\textbf{L}=\textbf{I} - \textbf{D}^{-1/2} \textbf{A} \textbf{D}^{-1/2} ,
\label{eqn:LaplacianMatrix}
\end{equation}
where $\textbf{I}$ denotes the identity matrix. 
Here, the normalized Laplacian matrix is used because it serves in the approximation of the minimization of $NCut$.

Next, we calculate the $k$ generalized eigenvectors from the normalized Laplacian matrix using the following notation:
\begin{equation}
\textbf{L}u = \lambda \textbf{D}u
\end{equation}
where vector $u$ is computed using the $k$ smallest eigenvalue.
Subsequently, the optimal $k$ value to use for clustering is computed using the Davies-Bouldin Index~\cite{davies1979cluster}.
After formulating the clustering process from the similarity metrics to clustering notation, The algorithm for sensor nodes profiling can be formulated as shown in Algorithm~\ref{alg:stayPointClustering}. 

\begin{algorithm}[h]	
	\caption{Sensors' Profiles Clustering Algorithm}
	\label{alg:stayPointClustering}
	\fontsize{8pt}{8pt}\selectfont
	\KwData{Valid data matrix, $\textbf{X}$}
	\KwResult{Cluster List, $C$}
	\BlankLine 
	1. Partition data into window $\textbf{X}'$ using Eqn.~\ref{eqn:windowsAveraging}. \BlankLine \vspace{-0.11cm}
    2. Calculate the normalized utilization similarity matrix, $\textbf{S}_{\mathtt{U}}$ as follows: \BlankLine \vspace{-0.11cm}
	\For{$i \leftarrow 1 $ to $n$}{
		\For{$j \leftarrow 1 $ to $n$}{
		 calculate the similarity metric using Eqn.~\ref{eqn:SimilarityMatrix}. \BlankLine \vspace{-0.11cm} 
		 calculate the distance using Eqn.~\ref{eqn:inverseDistance}. \BlankLine \vspace{-0.11cm} 
		}
	}
	3. Calculate the Temporal Features, $\textbf{S}_{\textbf{f}}$ as follows:  \BlankLine \vspace{-0.11cm} 
	\For{each time slot}{
		\For{$a \leftarrow 1 $ to $n$}{
			\For{$b \leftarrow 1 $ to $n$}{
				calculate the similarity using Eqn.~\ref{eqn:SimilarityMatrix}. \BlankLine \vspace{-0.11cm} 
				calculate the distance using Eqn.~\ref{eqn:inverseDistance}. \BlankLine \vspace{-0.11cm} 
			}
		}	
	}
	4. Calculate the affinity matrix, $\textbf{A}$ using Eqn.~\ref{eqn:affinityMatrixDerive}.\BlankLine \vspace{-0.11cm} 
	5. Compute the degree matrix, $\textbf{D}$ using Eqn.~\ref{eqn:affinityMatrixDerive}. \BlankLine \vspace{-0.11cm} 
	6. Calculate the Laplacian Matrix, $\textbf{L}$ using Eqn.~\ref{eqn:LaplacianMatrix}. \BlankLine \vspace{-0.11cm} 
	7. Calculate the Eigen Vector, $\textbf{U}$. \BlankLine \vspace{-0.11cm} 
	8. Determine $k$ value based on Davies-Bouldin Index based on $\textbf{U}$. \BlankLine \vspace{-0.11cm} 
	9. Perform  $k$-means and obtain cluster list, $C$. \BlankLine \vspace{-0.11cm} 
    10. return $C$.
\end{algorithm}

\begin{figure*}[b!]
	\includegraphics[width=0.911\textwidth]{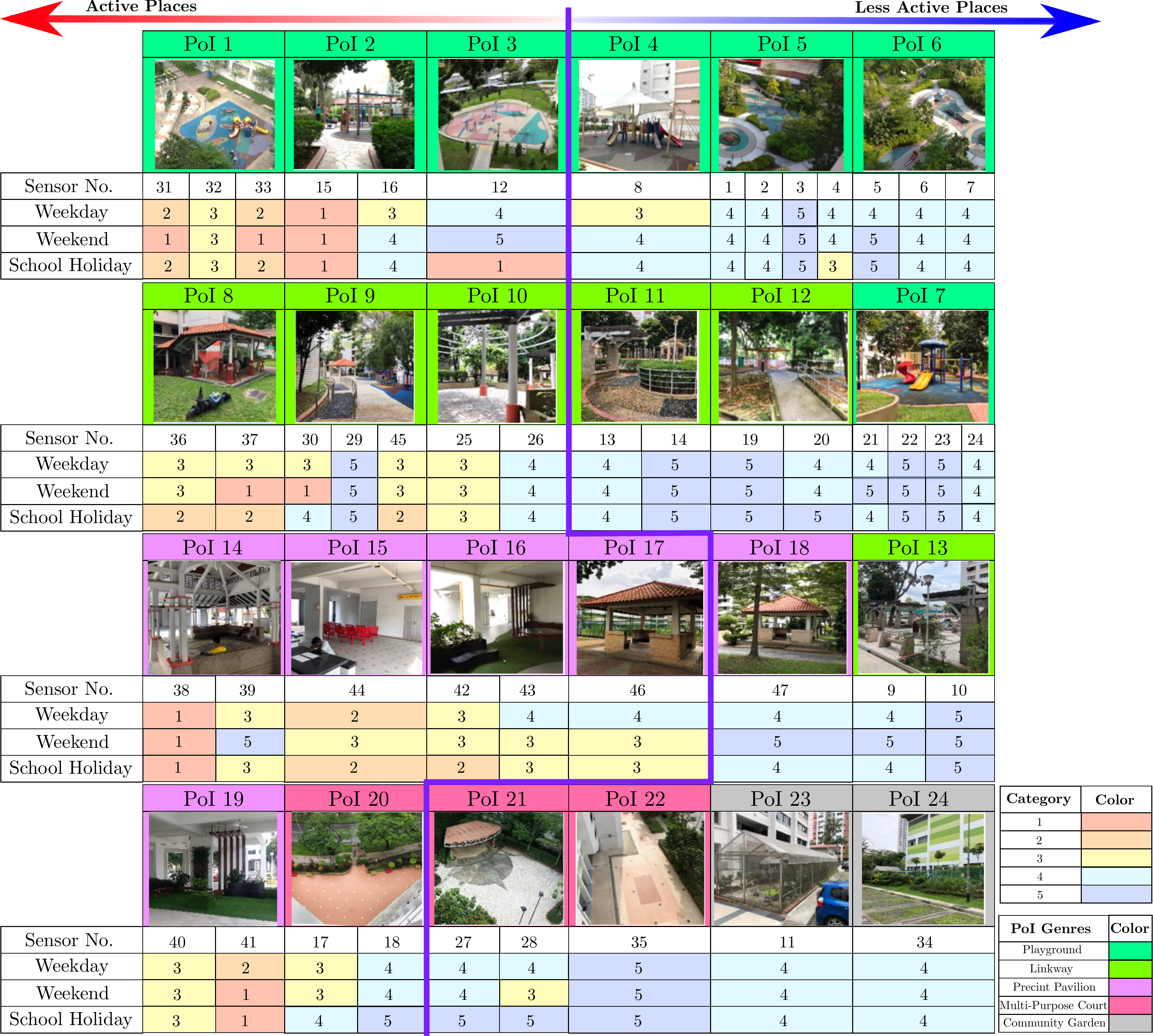} 
	\centering \caption{The separation of active and less active PoI based on the clusters (it needs at least two Category $3$ or more active cluster's category to be denoted as active).}
	\label{fig:ClusterPoiResult}
\end{figure*}

To this end, we generate different profiles based on large scale sensors w.r.t. temporal features. 
The challenges encountered in designing the data processing pipeline can be divided into the following three steps, which are (1) similarity study, (2) temporal elements, and (3) clustering approach. 
A similarity study between sensor nodes is necessary to determine whether data captured by sensors are similar. Among other conventional similarity studies, most of them (Euclidean, Manhanttan, and Minkowski) do not include temporal elements into similarity computation. 
Hence, WIED is used to calculate similarity metrics with consideration of data input over time using sliding windows.
In order to minimize the inclusion of midnight temporal, the temporal sessions are divided into four sessions. The main reason for performing such operations to have a higher weightage in computing similarity metrics between two sensors at the active period.
Meanwhile, the detail of the clustering is discussed in the next section.
Since an unsupervised machine learning approach is applied to profile the normalized utilization patterns into similar categories, some sort of empirical result interpretation is still required.
Thus, we will further explore the cluster results based on the profiling algorithm in the next section and study the underlying factors affecting public space utilization.

%%%%%%%%%%%%%%%%%%%%%%%%%%%%%%%%%%%%%%%%%%%%%%%%%%%%%%%%%%%%%%%%%%%%%%%%%%%%%%%%%%%%%%%%%%%%%%%%%%%%%%%%%%%%%%%%%%%%%%%%%%
\begin{figure*}[b!]
	\includegraphics[width=0.9\textwidth]{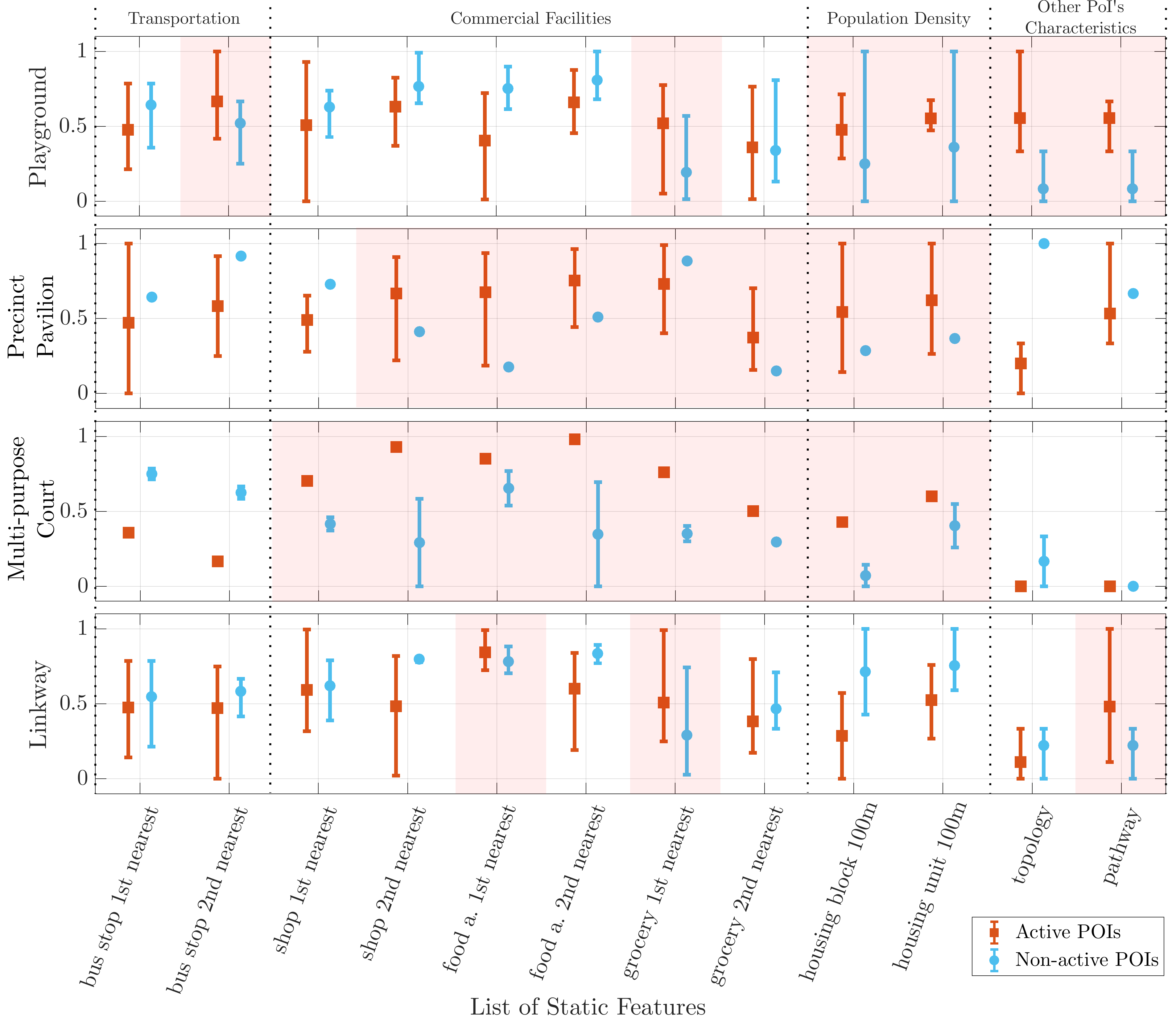} 
	\centering 
	\caption{Comparison of static features between active and less active PoIs (significant features are highlighted in red color)}
	\label{fig:static_features}
\end{figure*}

\section{Study of Utilization Profiles}
\label{sec:magnitudeUtz}
In this section, we use the proposed sensor profiling model from the previous session, and group PoI with similar normalized utilization patterns for generic temporal labels.
The nine temporal labels are merged into three generic temporal labels as shown in Table~\ref{tbl:TemporalLabelGenerate}, which are (1) weekday, (2) weekend, and (3) school holiday.
\begin{table}[H]
	\fontsize{7pt}{7pt}\selectfont
	\centering
	\caption{Temporal Labels Aggregation}
	\label{tbl:TemporalLabelGenerate}
\begin{tabular}{@{}c|c@{}}
\toprule
Generic Temporal Label & Temporal Label \\ \midrule
Weekday(WD) & \begin{tabular}[c]{@{}c@{}}Monday, Tuesday, Wednesday, \\ Thursday, Friday\end{tabular} \\
Weekend(WE) & Saturday, Sunday \\
School Holiday(SH) & School Holiday \\ \bottomrule
\end{tabular}
\end{table}
The extraction of the three generic temporal labels is computed using average function over daily data (288 samples) for each timestamp of the day over assigned temporal labels.
Note that public holidays are not included in the profiling stage as the number of samples collected is insufficient, and does not yield any interesting or meaningful findings due to lack of samples.

\subsection{Cluster Results on Utilization Profiles}
After preprocessing the raw data, the windowed data from an average function based on Eqn.~\ref{eqn:windowsAveraging} are shown in the following Fig.~\ref{fig:AverageUtz}(a). 
As observed from the windowed average data, we notice that the non-active period (12:00am-06:00am) has less public space utilization compared to the active period.
This causes some distortion in the distance matrix when calculating the similarity between the non-active period (roughly $6$ hours) among sensor nodes.
Hence, the proposed Algorithm~\ref{alg:stayPointClustering} is used to perform temporal session-based similarity measurement between sensor nodes by reducing the weights of the similarity matrix.
Next, based on the similarity matrix, we need to determine the optimal $k$ values to perform clustering.
Other than relying on Davies-Bouldin Index to choose the optimal $k$ value, the Eigen components are visualized to have a visual cue about the clustering process as illustrated in Fig.~\ref{fig:AverageUtz}(b).

%
%\label{fig:AverageUtz2
To investigate the different time of the week, we cluster the sensor nodes based on their utilization value, and the result is shown as follows:
\begin{enumerate*}[label={\alph*)}]
	\item Weekday - $5$ clusters
	\item Weekend - $4$ clusters, and
	\item School Holiday - $5$ clusters.
\end{enumerate*}
For each cluster, the average normalized utilization value is computed for each sensor node illustrated in Fig.~\ref{fig:AverageUtz}(a).
After that, the average normalized utilization value for each cluster data is inspected, and then define the utilization profile for each cluster. 
Weekday and school holiday's clusters yield similar $k$ value, while the weekend category has one cluster less in contrast. 
In order to further understand the underlying interpretation of each cluster, the average normalization utilization value of each sensor node within the cluster is calculated, which is also shown in Fig.~\ref{fig:temporalMatrix}.

Based on the result inspection, there are five different utilization patterns for weekdays and school holidays, while the weekend only has four utilization patterns.
The categories are sorted according to the normalized utilization average value from high to low, which indicates the Category $1$ is a high utilization group, while Category $5$ denotes low utilization.
The Category $4$ is composed of sensor nodes that have a rather low or short burst utilization, where Category $5$ sensor nodes have a rather low utilization data on the public space.
Sensor nodes in Category $3$ are somewhere in between Category $2$ and $4$, where there is some public space utilization, but much lower than Category $2$ and $1$.
According to Fig.~\ref{fig:temporalMatrix}, there are more sensor nodes labeled as Category $1$ during the weekend and school holidays.
During the school holiday, there are also more sensor nodes in Category $1$ and Category $2$ as compared to weekday and weekend. 
These phenomenons imply that PoIs are more active during the weekend and school holidays.
Interestingly, there are more sensor nodes that belong to Category $5$ during the weekend and school holiday compared to weekdays.

\subsection{Study of the Clusters' Static Features}
\label{secsub:staticfeaturesStudy}
To generalize the clustering result, we first divide the PoIs with respective sensor nodes into active and less active PoIs. 
Three different temporal characteristics are plotted side by side indicated by different colors in order to have an overview of PoI activeness.
The PoIs with at least two categories more active than Category $3$ are defined as a baseline to decide whether the specific PoI is considered as active or less active.
Subsequently, the cluster results for all three different temporal categories are tabulated as shown in Fig.~\ref{fig:ClusterPoiResult}, and observe different types of PoIs sorted according to their activeness.
After partitioning the PoI into active and inactive PoIs, we further explore the static features of each PoI through averaging according to the list. 
Since there are various kinds of static features available, and only factors that can be quantified are generated such as (1) transportation, (2) commercial facilities, (3) population density, and (4) other PoI characteristics.
Details of the static features can be found in Table~\ref{tbl:staticFeats}.

\begin{table}[h!]
	\fontsize{7pt}{7pt}\selectfont
	\centering
	\caption{Static Features for PoIs, $Z'$}
	\label{tbl:staticFeats}
	\begin{tabular}{c|l|l}
		\hline
		\textbf{\begin{tabular}[c]{@{}c@{}}Type of \\ Features\end{tabular}} & \multicolumn{1}{c|}{\textbf{Features Name}} & \multicolumn{1}{c}{\textbf{Description}} \\ \hline
		\multirow{2}{*}{Transportation} & \begin{tabular}[c]{@{}l@{}} 1st nearest bus stop\\ (distance feature in meters)\end{tabular}& \begin{tabular}[c]{@{}l@{}}Nearest bus stop \\ to a particular PoI \end{tabular} \\ \cline{2-3} 
		& \begin{tabular}[c]{@{}l@{}}2nd nearest bus stop\\ (distance feature in meters)\end{tabular} & \begin{tabular}[c]{@{}l@{}}Second nearest bus\\  stop to a particular PoI \end{tabular} \\ \hline
		\multirow{6}{*}{\begin{tabular}[c]{@{}c@{}}Commercial \\ Facilities\end{tabular}} & \begin{tabular}[c]{@{}l@{}}1st nearest shop \\(distance feature in meters)\end{tabular} & \begin{tabular}[c]{@{}l@{}}Nearest shop \\(non grocery and food amenity)\end{tabular} \\ \cline{2-3} 
		& \begin{tabular}[c]{@{}l@{}}2nd nearest shop \\ (distance feature in meters)\end{tabular}& \begin{tabular}[c]{@{}l@{}}Second nearest shop \\(non grocery and food amenity)\end{tabular} \\ \cline{2-3} 
		& \begin{tabular}[c]{@{}l@{}}1st nearest food amenity \\ (distance feature in meters) \end{tabular} & \begin{tabular}[c]{@{}l@{}}Nearest food amenity \\  (coffee house/hawker center) \end{tabular} \\ \cline{2-3} 
		& \begin{tabular}[c]{@{}l@{}}2nd nearest food amenity\\ (distance feature in meters)  \end{tabular} & \begin{tabular}[c]{@{}l@{}}Second nearest food amenity\\ (coffee house/hawker center) \end{tabular} \\ \cline{2-3} 
		& \begin{tabular}[c]{@{}l@{}}1st nearest grocery \\(distance feature in meters)\end{tabular}& \begin{tabular}[c]{@{}l@{}}Nearest grocery shop \\(Provision/wet market/kiosk) \end{tabular} \\ \cline{2-3} 
		& \begin{tabular}[c]{@{}l@{}}2nd nearest grocery\\ (distance feature in meters) \end{tabular}& \begin{tabular}[c]{@{}l@{}}Second nearest grocery shop \\ (Provision/wet market/kiosk) \end{tabular} \\ \hline
		\multirow{2}{*}{\begin{tabular}[c]{@{}c@{}}Population \\ Density\end{tabular}} & \begin{tabular}[c]{@{}l@{}}Number of \\Housing Block\end{tabular} & \begin{tabular}[c]{@{}l@{}	}Number of Housing \\ blocks within 100$m$\end{tabular} \\ \cline{2-3} 
		& \begin{tabular}[c]{@{}l@{}}Number of \\Housing Unit\end{tabular} & \begin{tabular}[c]{@{}l@{}}Number of Housing \\ unit within 100$m$\end{tabular} \\ \hline
		\multirow{2}{*}{\begin{tabular}[c]{@{}c@{}}Other PoI \\ Characteristics\end{tabular}} & Topology of PoI & \begin{tabular}[c]{@{}l@{}}Number of surrounding \\ building(separate			d by roads)\end{tabular} \\ \cline{2-3} 
		& \begin{tabular}[c]{@{}l@{}}Connected \\ Pathway\end{tabular} & \begin{tabular}[c]{@{}l@{}}Number of the connected\\ pathway within 50$m$\end{tabular} \\ \hline
	\end{tabular}
\end{table}

To draw a fair comparison between different static features, the min-max normalization method shown in Eqn.~\ref{eqn:minmax} is used for normalizing the static features for all the sensor inputs. 
In this context, we adopt two-extreme end empirical methods to compare the active and non-active PoIs. 
Each group will have their normalized static features value averaged according to the number of active and non-active PoIs.
The result is illustrated in Fig.~\ref{fig:static_features}, and the community garden are omitted as both of them are marked as inactive. 
For easier interpretability of the result, the significant result is highlighed in red color, where the average of the active PoIs is higher than the non-active PoIs.
Note that the average of static data is compuated and the distribution may have a wider range, which is harder to conclude.

Based on observation, the population density category is one of the prominent features that distinguish active and non-active playgrounds, in which the average of housing block and unit for active PoIs is higher than the non-active PoIs.
In addition, the pathway connection to the playground is possibly another reason why one playground is more active than another.
The majority of the precinct pavilions are affected by the neighboring facilities such as food amenities and grocery shops.
Similar to the precinct pavilion, multi-purpose court PoI is affected by the neighboring facilities with different kinds of shops being considered important features.
Besides, the link-way does not seem to have a distinct pattern on the static features that dictate its activeness. 
This may due to the limited coverage of the sensor nodes causes some potential link-ways that are not captured, and therefore the trajectory of residents might not be captured in detail.
And lastly, we observed that majority of the PoI locations are not affected by the accessibility of public transport.

In a nutshell, public space utilization profiles are highly dependent on their static spatial features.
One of the significant features that affect the public space utilization based on the observation is the population density of a surrounding PoI, where a naturally higher population implies higher public space utilization. 
Also, the availability of nearby commercial facilities plays an important role in ensuring the liveliness of a PoI.
Such observations are also aligned with the findings from social behavior study that is conducted in the survey found in~\cite{Chong2019Role} and~\cite{Chong2019When}.

%%%%%%%%%%%%%%%%%%%%%%%%%%%%%%%%%%%%%%%%%%%%%%%%%%%%%%%%%%%%%%%%%%%%%%%%%%%%%%%%%%%%%%%%%%%%%%%%%%%%%%%%%%
\section{Conclusion}
\label{sec:conclusion}
%%%%%%%%%%%%%%%%%%%%%%%%%%%%%%%%%%%%%%%%%%%%%%%%%%%%%%%%%%%%%%%%%%%%%%%%%%%%%%%%%%%%%%%%%%%%%%%%%%%%%%%%%%
We present a space-centric system model for processing data collected from public space in an urban residential area. 
The potential factors that drive public space utilization are identified through the correlation of the IoT data sources, which are based on static features of the region studied.
This allows us to have an in-depth understanding of public space utilization using heuristic correlation.
We would like to draw a few conclusions based on analysis of the result as follows:

\begin{itemize}
\item The activeness of each PoI over a region over weekday, weekend, and school holiday is identified.
\item Using the spectral clustering method, space-centric sensors can be segmented into different levels of activeness.
\item From the public space utilization profile's result, the liveliness of a PoI can be highly correlated with the static features such as nearby commercial facilities and population density.
\end{itemize}

However, some questions remain such as how the identified factors can be used to build a more livable area or guide the urban designer to better plan for future development.
It should be noted that this is a case study of a specific region, and it may not be representative across the nationwide or different country cultures.
Also, should the number of sensors coverage increase, a different result can be obtained as different residential areas may have varying behavior and diversifying factors. 
Those questions will be included in future works to further generalize factors and design models to aid urban planners to better plan the residential area development.

\section*{Acknowledgment}
%%%%%%%%%%%%%%%%%%%%%%%%%%%%%%%%%%%%%%%%%%%%%%%%%%%%%%%%%%%%%%%%%%%%%%%%%%%%%%%%%%%%%%%%%%%%%%%%%%%%%%%%%%
This work is supported by Singapore Ministry of National Development (MND) Sustainable Urban Living Program, under the grant no. SUL2013-5, ``Liveable Places: A Building Environment Modeling Approach for Dynamic Place Making” project, and especially appreciate the useful discussion and help from the collaborators from MND, HDB and URA.

\bibliographystyle{IEEEtran}
\newcommand{\BIBdecl}{\setlength{\itemsep}{0.25 em}}
\bibliography{bibSpace}

\end{document}